\documentclass[aps,prc,reprint,twocolumn,superscriptaddress,nofootinbib]{revtex4-2}

\usepackage{hyperref}
\usepackage{mathtools}
\usepackage{amsmath}
\usepackage{amssymb}
\usepackage{amsbsy}
\usepackage{bm}
\usepackage{textcomp}
\usepackage[dvips]{graphicx,color}
\usepackage{chngcntr}
\usepackage{tikz}
\usetikzlibrary{shapes.geometric, arrows}
\usepackage{fancyhdr}
\usepackage{float}
\usepackage{dsfont}

\newcommand{\sixj}[6]{\left\{\begin{array}{ccc} #1 & #2 & #3 \\ #4 & #5 & #6 \\ \end{array}\right\}}

\newcommand{\half}{\frac{1}{2}}

\newcommand{\lla}{\left\langle}
\newcommand{\rra}{\right\rangle}

\newcommand{\Nmax}{$N_{\rm max}$}

\newcommand{\braketop}[3]{\ensuremath{\left\langle #1 \right| #2 \left| #3 \right\rangle}}

\newcommand{\RedME}[3]{\ensuremath{\langle #1 \| #2 \| #3 \rangle}}

\newcommand{\bra}[1]{\ensuremath{\left\langle #1 \right|}}
\newcommand{\ket}[1]{\ensuremath{\left| #1 \right\rangle}}
\newcommand{\braket}[2]{\ensuremath{\langle #1 | #2 \rangle}}

\newcommand{\hw}{\ensuremath{\hbar\Omega}}

\newcommand{\iu}{\mathrm{i}\mkern1mu}

\begin{document}

\title{\textit{Ab initio} translationally invariant nucleon-nucleus optical potentials
}
\author{M. Burrows}
\affiliation{Department of Physics and Astronomy, Louisiana State University, Baton Rouge, LA 70803, USA}
\author{K. D. Launey}
\affiliation{Department of Physics and Astronomy, Louisiana State University, Baton Rouge, LA 70803, USA}
\author{A. Mercenne}
\affiliation{Department of Physics and Astronomy, Louisiana State University, Baton Rouge, LA 70803, USA}
\author{R. B. Baker}
\affiliation{Institute of Nuclear and Particle Physics, and Department of Physics and Astronomy, Ohio University, Athens, OH 45701, USA}
\author{G. H. Sargsyan}
\affiliation{Lawrence Livermore National Laboratory, Livermore, CA 94550, USA}
\author{T. Dytrych}
\affiliation{Nuclear Physics Institute, Academy of Sciences of the Czech Republic, 250 68 Řež, Czech Republic}
\affiliation{Department of Physics and Astronomy, Louisiana State University, Baton Rouge, LA 70803, USA}
\author{D. Langr}
\affiliation{Department of Computer Systems, Faculty of Information Technology, Czech Technical University in Prague, Prague 16000, Czech Republic}

\begin{abstract}
We combine the \textit{ab initio} symmetry-adapted no-core shell model (SA-NCSM) with the single-particle Green's function approach to construct optical potentials rooted in first principles. Specifically, we show that total cross sections and phase shifts for neutron elastic scattering from a $^4$He target with projectile energies between 0.5 and 10 MeV closely reproduce the experiment. 
In addition, we discuss an important new development that resolves a long-standing issue with spurious center-of-mass motion in the Green's function formalism for many-body approaches. The new development opens the path for first-principle predictions of cross sections for elastic scattering of single-nucleon projectiles, nucleon capture and deuteron breakup reactions, feasible for a broad range of open-shell spherical and deformed nuclei
in the SA-NCSM approach.

\end{abstract}

\maketitle

\section{Introduction}

Remarkable progress has been made in recent years in the development of many-body approaches from first principles to scattering and nuclear reactions (see Refs. \cite{FRIBTAwhite2018,1402-4896-91-5-053002,0954-3899-41-12-123002} for  reviews), 
including, e.g.,  studies of elastic scattering \cite{NollettPWCH07,HagenDHP07,PhysRevLett.101.092501,ElhatisariLRE15,QuaglioniN09,BurrowsBEWLMP20,Mercenne:2021hef}, photoabsorption \cite{PhysRevC.90.064619}, transfer \cite{NavratilQ12}, neutron capture reactions \cite{PhysRevLett.105.232502}, as well as resonant states \cite{lazauskas_2019}, thermonuclear fusion \cite{HupinQN19} and alpha capture reactions \cite{DreyfussLESBDD20}.

A more general approach to reactions, especially suitable for heavier nuclei, is based on identifying few-body degrees, typically the  reaction fragments (or clusters) involved in the reaction, and reducing the many-body problem to a few-body technique \cite{ThompsonN09}. As this reduction results in  effective interactions (often referred to as optical potentials) between the clusters, the critical need for parameter-free  interactions has been recognized as one moves away from stability, where uncertainties become uncontrolled since elastic scattering data does not uniquely constrain the optical potential  \cite{LovellN15}.

To address this, recent studies have utilized realistic inter-nucleon interactions, typically derived in the chiral effective field theory, without the need to fit interaction parameters in the nuclear medium. These models have built upon earlier theoretical frameworks, such as the one  introduced by Feshbach, leading to the Green's function formulation \cite{capuzzi2000223} and to the successful dispersive optical model \cite{Mahaux:1986zz,Mahaux1991,DickhoffBook}, as well as the one pioneered by Watson~\cite{Watson1953a,KMT} for elastic scattering of a single-nucleon projectile, leading to the spectator expansion of the multiple scattering theory~\cite{Siciliano:1977zz}. Successful recent applications include \textit{ab initio} nucleon-nucleus potentials for elastic scattering for closed-shell nuclei at low projectile energies ($\lesssim20$ MeV per nucleon) based on the Green's function technique with the coupled-cluster method \cite{RotureauDHNP17,PhysRevC.98.044625}  and the self-consistent Green's function method \cite{idini19}, as well as for light targets in the intermediate-energy regime ($\gtrsim 65$ MeV per nucleon) using the spectator expansion of the multiple scattering theory and the \textit{ab initio} no-core shell model  \cite{BurrowsBEWLMP20,VorabbiGFGNM2022} (see also Ref. \cite{FRIBTAwhite2018}). Similarly, optical potentials have been derived from two- and three-nucleon chiral forces in nuclear matter \cite{WhiteheadLH2019}. These potentials provide cross sections for elastic proton or neutron scattering, and in addition can be used as input to modeling (d,p) and (d,n) reactions \cite{Rotureau_2020}.

In this paper, we construct \textit{ab initio} nucleon-nucleus (NA) optical potentials that are translationally invariant (t.i.) and applicable to a broad range of open-shell spherical and deformed nuclei. We achieve this by combining the Green's function (GF) approach with the symmetry-adapted no-core shell model (SA-NCSM) \cite{DytrychLDRWRBB20,LauneyMD_ARNPS21}, which accommodates from single-particle features to collective and clustering correlations in nuclei. In addition, an important advantage of the GF technique is that the NA effective potentials include the information about all near reaction channels  through the GF calculations in the $(A\pm1)$ systems. In this SA-NCSM/GF framework, we illustrate the new developments for the elastic neutron scattering off the $^4$He ground state, and show that phase shifts and cross sections agree remarkably well with experimental values. 
This reaction has been previously studied at length in many-body calculations without explicitly constructing optical potentials, including, e.g., the no-core shell model with continuum (NCSMC) \cite{QuaglioniN09,Kravvaris:2020lhp}, Faddeev-Yakubovsky approach \cite{Lazauskas2018}, and the single-state harmonic oscillator representation of scattering equations \cite{PhysRevC.98.044624}, and hence provides a well-informed case for theoretical benchmarks.

An important feature of the SA-NCSM/GF optical potentials is that they are translationally invariant. Specifically, it has been long recognized that the translational invariance is violated in Green's functions calculated within a many-body framework that uses laboratory coordinates and a single-particle mean-field basis, in which the nucleon antisymmetrization is fully taken into account using configuration representation but introduces center-of-mass (CM) spuriosity (see, e.g., Refs.~
\cite{Johnson:2017xdg,Johnson:2019qbe}). In earlier studies of heavy target systems the CM spuriosity has been neglected  due to its $A$ dependence. For \textit{ab initio} calculations, especially for light targets or in the case of the target and projectile having similar masses, ensuring the translational invariance is critical. 
To achieve this, in this study, we utilize the Lawson procedure \cite{Lawson74} that has been successfully used in many-body nuclear structure calculations. We emphasize that in complete no-core shell-model (NCSM) spaces (truncated by the total number of harmonic oscillator excitations) \cite{Barrettnv13} and in selected model spaces of the SA-NCSM, the center of mass wavefunction can be factored out exactly \cite{VERHAAR1960508,Hecht:1971xlg,Millener92,launeydd16}, leading to an exact removal of the CM spuriosity in the Lawson procedure (see Sec. \ref{tiGFfromLab} for details). In addition, the SA-NCSM provides a correct treatment of collective and cluster correlations, including coupling to continuum degrees of freedom (dof), which makes the framework especially suitable to accommodate these effects in calculations of cross sections and in studies of absorption, target deformation, and low-lying resonances.
The new developments provide a tool for first-principle predictions of cross sections for elastic neutron and proton scattering, as well as for constructing NA optical potentials for neutron/proton capture, (d,p) and (d,n) reactions.

\section{Theoretical Framework}

For completeness, we briefly outline the Green's function theory and its relation to nucleon-nucleus optical potentials, as introduced and reviewed in earlier papers \cite{capuzzi2000223,capuzzi1996147,PhysRevC.66.034313,Mahaux1991,DickhoffBook}. For a many-body system, the single-particle (s.p.)  time-ordered Green's function is defined as (see Ref. \cite{PhysRevC.66.034313})
\begin{widetext}
\begin{eqnarray}
	G( \mathbf{r},\mathbf{r}';E) = \lim_{\epsilon \rightarrow 0} \bra{\Psi_0^A} [a_{\mathbf{r}}+a_{\mathbf{r}'}^\dagger] \frac{1}{E - (\hat H - E^A_0 - \iu\epsilon)(\hat N-A)} [a_{\mathbf{r}}+a_{\mathbf{r}'}^\dagger] \ket{\Psi_0^A},
 \label{GFdef}
\end{eqnarray}
\end{widetext}
where $\hat H$ is the many-body realistic Hamiltonian, $E$ is the energy in the center-of-mass frame, 
$\ket{\Psi_0^A}$ is the ground state (g.s.) of the $A$-body target nucleus with energy $E_0^A$ and total angular momentum $J_0$ (or any given target state of interest), and $a^{(\dagger)}_{\mathbf{r}}$ annihilate (create) a particle at position $\mathbf{r}$ relative to the center of mass of the target
(we note that, for simplicity of notations  $\lim_{\epsilon \rightarrow 0}$ will be omitted but implied for
all further Green's function equations).
The operator $\hat N= \int d \mathbf{r} a_{\mathbf{r}}^\dagger a_{\mathbf{r}} $ is the particle number operator, which commutes with the Hamiltonian, and yields the eigenvalues $\hat N \ket{\Psi^{A\pm 1}}=(A\pm1) \ket{\Psi^{A\pm 1}}$. The operator $\hat{G}(E,\epsilon)$ is then defined as:
\begin{equation}
 \hat{G}(E,\epsilon) = \frac{1}{E - (\hat H - E^A_0 - \iu\epsilon)(\hat N-A)}.
 \label{Gop}
\end{equation}
The Green's function is calculated in an orthonormal basis, $\ket{\Phi}=[a+a^\dagger] \ket{\Psi_0^A}$ with a norm
\begin{equation} 
	\mathcal{N}({\mathbf{r}},{\mathbf{r}}')=\mathcal{N}^p({\mathbf{r}},{\mathbf{r}}')+\mathcal{N}^h({\mathbf{r}}',{\mathbf{r}})=\delta(\mathbf{r}-\mathbf{r}'),
\label{Eq:Norm}
\end{equation}
where $\mathcal{N}^p({\mathbf{r}},{\mathbf{r}}')=\bra{\Psi_0^A} a_{\mathbf{r}}a_{\mathbf{r}'}^\dagger \ket{\Psi_0^A}$ is the norm of the particle states
and $\mathcal{N}^h({\mathbf{r}'},{\mathbf{r}})=\bra{\Psi_0^A} a_{\mathbf{r}'}^\dagger a_{\mathbf{r}} \ket{\Psi_0^A} \equiv \rho({\mathbf{r}}',{\mathbf{r}})$ is the norm of the hole states, equivalent to the one-body density $\rho({\mathbf{r}}',{\mathbf{r}})$ of the target state.

The equation of motion (EoM) for the s.p. propagator (\ref{GFdef}) is \cite{PhysRevC.66.034313,Mahaux1991}:
\begin{eqnarray}
    &&(E-T_{\rm rel}(\mathbf{r}))G(\mathbf{r},\mathbf{r'};E)-\int d \mathbf{r}''V(\mathbf{r},\mathbf{r''};E)G(\mathbf{r''},\mathbf{r'};E) \cr
    &&=\delta (\mathbf{r}-\mathbf{r'}),
    \label{eq:EoMGF}
\end{eqnarray}
where $T_{\rm rel}$ is the relative kinetic energy in the CM frame of the two-cluster system. $V(\mathbf{r},\mathbf{r''};E)$  describes the interaction of the propagating particle or hole with all the other particles or holes in the medium at energy $E$. For energies above the single-nucleon threshold, this provides the effective interaction between the single-nucleon projectile and target, and will be referred to as nucleon-nucleus optical potential. Using the EoM (\ref{eq:EoMGF}), $V(\mathbf{r},\mathbf{r'})$ can be calculated as
\begin{equation}
    V(\mathbf{r},\mathbf{r'};E)=(E-T_{\rm rel}(\mathbf{r}))\delta (\mathbf{r}-\mathbf{r'})-G^{-1}(\mathbf{r},\mathbf{r'};E),
    \label{eq:Vcoord}
\end{equation}
which is nonlocal and depends on $E$.
 
Eq.~(\ref{GFdef}) can be written in configuration representation 
using s.p. wavefunctions $\phi_{am_a}(\mathbf{r})$, for which $G( \mathbf{r},\mathbf{r}';E)=\sum_{\alpha m_\alpha \beta m_\beta} \phi_{\alpha m_\alpha}(\mathbf{r})\phi^*_{\beta m_\beta}(\mathbf{r}') G_{\alpha m_\alpha \beta m_\beta}(E)$, where
$a_\mathbf{r}=\sum_{\alpha m_\alpha} \phi_{\alpha m_\alpha}(\mathbf{r})a_{\alpha m_\alpha}$ and $a^\dagger_\mathbf{r}=\sum_{\alpha m_\alpha} \phi^*_{\alpha m_\alpha}(\mathbf{r})a^\dagger_{\alpha m_\alpha}$. In this study we use harmonic oscillator (HO) single-particle wavefunctions with $\alpha \equiv \{ n_\alpha (\ell_\alpha \half) j_\alpha \}$ and $\beta$ being the HO quantum numbers associated with the projectile ($n=2n_r+\ell$ is the HO shell number and $n_r$ is the radial quantum number). The $a_{\alpha m_\alpha}=(a_{\alpha m_\alpha}^\dagger)^\dagger$ operators are the usual annihilation and creation operators of a particle in a HO single-particle state $\ket{\alpha m_\alpha}$.
It is clear that for $J_0=0$ (cf. Ref. \cite{RotureauDHNP17}), Eq.~(\ref{GFdef}) yields
\begin{widetext}
\begin{eqnarray}
	G_{\alpha m_\alpha \beta m_\beta}(E) &=& \lla \Psi_0^A \left| a_{\alpha m_\alpha} \frac{1}{E - (H - E^A_0) + \iu\epsilon} a^{\dagger}_{\beta m_\beta} \right| \Psi_0^A \rra + \lla \Psi_0^A \left| a^{\dagger}_{\beta m_\beta} \frac{1}{E - (E^A_0- H) - \iu\epsilon} a_{\alpha m_\alpha} \right| \Psi_0^A \rra , 
 \label{GFconfig}
\end{eqnarray}
\end{widetext}
where one can define
particle and hole states:
\begin{eqnarray}
	\ket{ \Phi^+_{\alpha m_\alpha;J_0(M_0)} } \equiv a_{\alpha m_\alpha}^{\dagger} \ket{\Psi_{0,J_0M_0}^A}, \cr
	\ket{ \Phi^-_{\alpha m_\alpha;J_0(M_0)} } \equiv  a_{\alpha m_\alpha} \ket{ \Psi_{0,J_0M_0}^A}.
\end{eqnarray}

Equivalently, the SA-NCSM uses SU(3) proper tensors, $a_{(n_\alpha\,0) l_\alpha j_\alpha m_\alpha}^\dagger \equiv a_{\alpha m_\alpha}^\dagger$ and $\tilde a_{(0\,n_\alpha) l_\alpha j_\alpha -m_\alpha} =(-)^{n_\alpha+j_\alpha-m_\alpha} a_{\alpha m_\alpha} $, and cluster basis states with good total angular momentum:
\begin{widetext}
\begin{eqnarray}
	\ket{ \Phi^{J(M)+}_{ J_0\alpha} } \equiv (-1)^{j_\alpha+J_0-J}\left\{ a_{\alpha}^{\dagger} \times \ket{\Psi_{0,J_0}^A} \right \}^{J(M)}
	=\sum_{t}\frac{(-1)}{\Pi_{J}}\ket{tJ(M)}\RedME{tJ}{a_{\alpha}^{\dagger}}{\Psi_{0,J_0}^A}, \cr
		\ket{ \Phi^{J(M)-}_{ J_0 \alpha }} \equiv (-1)^{n_\alpha}(-1)^{j_\alpha+J_0-J}\left\{ \tilde a_{\alpha} \times \ket{\Psi_{0,J_0}^A} \right \}^{J(M)}
	=\sum_{t}\frac{(-1)^{1+n_\alpha}}{\Pi_{J}}\ket{tJ(M)}\RedME{tJ}{\tilde a_{\alpha}}{\Psi_{0,J_0}^A},
\label{Eq:pivots}
\end{eqnarray}
where $t$ is the complete many-body $A\pm 1$ basis and $\Pi_J = \sqrt{2J+1}$. 
The eigenfunctions $\ket{\Psi_{0,J_0}^A}$ are calculated in the SA-NCSM (or any many-body approach), whereas the basis vectors for each $\alpha$ and $J$, $\ket{ \Phi^{J \pm}_{ J_0\alpha} }$,  are calculated through the single-particle overlaps (since results do not depend on the $M$ projection, it is omitted from the notations).
In general, for a given $J_0=0$:
\begin{eqnarray}
	G^J_{J_0;\alpha\beta}(E) = \bra{\Phi^{J+}_{ J_0 \alpha }} \frac{1}{E - (H - E^A_0) + \iu\epsilon} \ket{\Phi^{J+}_{J_0 \beta } }  + \bra{\Phi^{J-}_{J_0 \beta }}  \frac{1}{E - (E^A_0- H) - \iu\epsilon}\ket{\Phi^{J-}_{J_0 \alpha }} \equiv G^{J+}_{J_0;\alpha\beta}(E) + G^{J-}_{J_0;\beta\alpha}(E),
 \label{GFconfig2}
\end{eqnarray}
which for $J_0=0$ coincides with Eq. (\ref{GFconfig}).
 
For the s.p. HO wavefunctions $\phi_\alpha(\mathbf{r})=R_{n_\alpha \ell_\alpha}(r)\mathcal{Y}_{(\ell_\alpha \half)j_\alpha}(\hat{r})=\sum_{m_\alpha\sigma_\alpha} C_{\ell_\alpha m_\alpha \half \sigma_\alpha}^{j_\alpha m} R_{n_\alpha \ell_\alpha}(r)Y_{\ell_\alpha m_\alpha}(\hat r)\chi_{\half \sigma_\alpha}$, with radial wavefunctions $R_{nl}(r)$ that are defined positive at infinity and spin functions $\chi_{\half \sigma}$, we obtain for $G$ (or  $G^{-1}$):
\begin{eqnarray}
    G_{J_0;\ell j \ell' j'}^{J}(r,r';E) &=&\sum_{M_0 m M_0' m'}\int d\hat r d\hat r' C_{J_0 M_0 j m}^{JM} \mathcal{Y}^\dagger_{(\ell\half)jm}(\hat r) G_{J_0M_0M_0'}(\mathbf{r},\mathbf{r}';E) \mathcal{Y}_{(\ell'\half)j'm'}(\hat r')C_{J_0 M_0' j' m'}^{JM} \cr
    &=&\sum_{n_\alpha n_\beta}R_{n_\alpha \ell}(r) R_{n_\beta \ell'}(r')G_{J_0;n_\alpha \ell j, n_\beta \ell' j'}^J(E) .
\end{eqnarray}
\end{widetext} 
The effective potential for the channels $\nu \equiv \{J_0;\ell j \}$ and $\nu'\equiv\{J_0;\ell' j'\}$ (or $\nu=\nu'=\{0^+;\ell \}$ for a $0^+$ target state) is then given by Eq. (\ref{eq:Vcoord}) as
\begin{eqnarray}
    V_{\nu\nu'}^J(r,r')&=&V_{J_0;\ell j \ell' j'}^{J}(r,r') \cr
    &=&(E-T_{\rm rel}(r))\delta_{\ell \ell'}\delta_{j j'}\frac{\delta(r-r')}{r r'} \cr
    &-&(G_{J_0\ell j \ell' j'}^{J})^{-1}(r,r',E).
\end{eqnarray}
Using that $T_{\rm rel}(r)\frac{\delta(r-r')}{r r'}$=$\sum_{n n'}^\infty R_{n \ell}(r) R_{n' \ell}(r') \bra{n\ell}T_{\rm rel}\ket{n' \ell}  $, we calculate $V$ for a finite $n_{\rm max}$ (similarly to the work of Ref. \cite{QuaglioniN09})
\begin{eqnarray}
    && V_{J_0;\ell j \ell' j'}^{J}(r,r') \cr
    &=& \delta_{\ell \ell'}\delta_{j j'} \sum_{n n'}^{n_{\rm max}} \left(
E\delta_{n n'}-\bra{n\ell}\hat T_{\rm rel}\ket{n' \ell} 
\right) R_{n \ell}(r) R_{n' \ell}(r') \cr
&-&\sum_{n n'}^{n_{\rm max}} 
(G_{J_0;n \ell j, n' \ell' j'}^J)^{-1}
  R_{n \ell}(r) R_{n' \ell'}(r') ,
    \label{optpot}
\end{eqnarray}
where $n_{\rm max}$ is the highest HO shell available to the $A$ and $A\pm 1$ systems, and is determined from \Nmax~ used in the SA-NCSM calculations (\Nmax~ is the total HO excitations above the nuclear configuration of the lowest HO energy). For the HO single-particle basis with radial wavefunctions that are positive at infinity, $\bra{n'\ell}T_{\rm rel}\ket{n \ell} = \frac{\hw}{2}\times $ 
\scalebox{.92}[1.0]{
$ \left( (n+\frac32) \delta_{n' n}-\sqrt{\frac{n-\ell}{2}\frac{n+\ell+1}{2}}\delta_{n' n-2}-\sqrt{\frac{n-\ell+2}{2}\frac{n+\ell+3}{2}}\delta_{n'n+2} \right).$
}
This ensures that at long distances the potential becomes zero. Calculations of the effective potential require an inversion of the Green's function, which we perform in configuration representation. Specifically, for given channels $\nu$ and $\nu'$, Eq. (\ref{optpot}) can be written as $\mathbf{V}=\mathbf{G}_0^{-1}-\mathbf{G}^{-1}$, where $\mathbf{G}_0^{-1} \equiv (E \mathds{1} - \mathbf{T}_{\rm rel}) $ is the free propagator for the two-cluster system, $\mathds{1} $ is the identity matrix, and  
$\mathbf{V}$ is a finite matrix with rows and columns enumerated by the radial quantum number $n_r=0,1,\dots,n_r^{\max}$ with the corresponding shell number $n\le n_{\rm max}$.

The optical potential is nonlocal and can enter as input to various reaction few-body approaches. In this study, we provide phase shifts evaluated in the  $\mathbf{R}$-matrix method \cite{Descouvemont_2010} with the SA-NCSM/GF $V(r,r')$ potential (\ref{optpot}). The $\mathbf{R}$-matrix method uses exact Coulomb eigenfunctions in the exterior region, whereas the Schr\"odinger equation holds in the interior region (with no Coulomb potential for neutron projectiles):
\begin{equation}
    (-E+T_{\rm rel}(r))\frac{u_\nu^J(r)}{r}+\sum_{\nu'}\int dr' r'^2 V_{\nu\nu'}^J(r,r')\frac{u_{\nu'}^J(r')}{r'}=0,
\end{equation}
where $u_\nu^J(r)$ are the wavefunctions of the relative motion of the projectile-target system up to a norm and can describe bound, resonance, and scattering states (see Appendix \ref{section:Numerical} for further discussions of the single-nucleon equation of motion).

\subsection{Calculations of a translationally invariant Green's function using laboratory coordinates
\label{tiGFfromLab}
} 

In this paper, we develop a method to calculate the translationally invariant (t.i.) Green's function derived in a many-body approach that uses laboratory coordinates. There has long been a problem with resolving the spurious center-of-mass (CM) contamination of Green's function calculations when using laboratory coordinates and configuration representations (or second quantization)~\cite{Johnson:2017xdg,Johnson:2019qbe}. Due to the decrease of the CM spurious effects with larger masses $A$, these effects have been neglected for heavy target systems and valence-shell calculations. 
The problem is that even if one works in a laboratory frame with the origin in the CM of the $A+1$ projectile-target system, the target and most importantly the second term in Eq. (\ref{GFconfig2}), which describes the hole states (or the $A-1$ system), necessarily have CM motion.
Using a transformation to Jacobi coordinates, we provide a mathematical construction that addresses the target CM motion. However, the particle and hole states in the Green's function, the first and second terms of  Eq. (\ref{GFconfig2}), respectively, need to be treated separately and 
with special care. To achieve this, in this study, we utilize the Lawson procedure \cite{Lawson74}, the same one used in many-body nuclear structure calculations, as outlined below.

In this study, calculations are carried through the SA-NCSM, where the single-particle Green's function is calculated in the laboratory frame. 
The nuclear Hamiltonian utilized in the SA-NCSM is non-relativistic and uses t.i. inter-nucleon interactions. The use of laboratory coordinates results in spurious center-of-mass excitation states, which are eliminated from the low-lying energy spectrum in structure calculations by using a Lawson term \cite{Lawson74}. 

The Lawson procedure utilizes a Lagrange multiplier term that is added to the intrinsic Hamiltonian expressed in laboratory coordinates, $H+\lambda_{\rm CM}\hat N_{\rm CM}$, where $\hat N_{\rm CM}$ is the operator that counts the number of CM excitations and $N^{\rm CM}$ is its eigenvalue that labels the CM component of the wavefunctions. For a typical value of $\lambda_{\rm CM} \sim 50$ MeV, the nuclear states of interest (with energy  $\lesssim 30$ MeV) have wavefunctions that are free of center-of-mass excitations ($N^{\rm CM}=0$), while CM-spurious states ($N^{\rm CM} > 0$) lie much higher in energy. 
It is important that in the conventional NCSM with complete model spaces truncated by $N_{\rm max}$ (see the review \cite{Barrettnv13}) and in selected model spaces of the SA-NCSM (see the review \cite{launeydd16}), the center-of-mass wavefunction can be factored out exactly. The reason is that the CM operator ($\hat N_{\rm CM}$) does not mix CM states with different HO excitations, and in addition, being an SU(3) scalar $(0\, 0)$, it does not mix SU(3) subspaces of the SA-NCSM \cite{VERHAAR1960508,Hecht:1971xlg,Millener92}. Hence,
each $A$-body wavefunction can be \emph{exactly } factorized to an intrinsic wavefunction that can be  equivalently expressed through Jacobi coordinates and a HO wavefunction of the CM with $\Gamma=\{N^{\rm CM}, L^{\rm CM}, M^{\rm CM}\}$, such that $\Psi^{A}_{\rm intr}(\pmb{\xi}_1 ,\dots, \pmb{\xi}_{A-1})\phi_{\Gamma}(\mathbf{R}_{\rm CM}^A)$ (see Sec. \ref{sec:Jacobi}; cf. \cite{Navratil:2004tw}), where the states with $N^{\rm CM}=0$ are the physical states of interest. Finally, since the Hamiltonian is translationally invariant, there is no contribution from the CM component to the matrix elements of the Hamiltonian or any function of the Hamiltonian, as in the case of the Green's function.

We derive the Green's function by using the completeness of the many-body Hamiltonian eigenfunctions for the $A\pm 1$ systems in Eq. (\ref{GFconfig2}), the so-called Lehmann representation, which for the laboratory frame (denoted as ``L") 
and for $J_0=0$ 
is\footnote{
Since $J_0$ is fixed from the reaction entrance channel and is the same for all calculations, we omit $J_0$ from the notations henceforth.
}:
\begin{widetext}
\begin{eqnarray}
	G^{J+}_{ab (\rm L)}(E) = 
 \sum_{k \Gamma_k} \frac{ \braket{\Phi^{J+}_a}{\Psi_{k\Gamma_k}^{A+1}}_{\rm L} \braket{\Psi_{k\Gamma_k}^{A+1}}{\Phi^{J+}_b}_{\rm L}  }{E - (\varepsilon_k^+ + \lambda_{\rm CM} N_{k}^{\rm CM}) + \iu\epsilon},\,
 G^{J-}_{ba (\rm L)}(E) = \sum_{k \Gamma_k}\frac{ \braket{\Phi^{J-}_b}{\Psi_{k\Gamma_k}^{A-1}}_{\rm L} \braket{\Psi_{k\Gamma_k}^{A-1}}{\Phi^{J-}_a}_{\rm L}  }{E - (\varepsilon_k^- - \lambda_{\rm CM} N_{k}^{\rm CM} ) - \iu\epsilon},
 \label{GFcompl}
\end{eqnarray}
\end{widetext} 
where $\varepsilon_k^+ \equiv E_k^{A+1}-E_0^A$, $\varepsilon_k^- \equiv E_0^A- E_k^{A-1}$,  the overlaps $\braket{\Phi^{J\pm}}{\Psi^{A\pm1}_{k\Gamma_k}}_{\rm L}$ are calculated in the laboratory frame using Eq. (\ref{Eq:pivots}), and the $A\pm 1$ eigenstates $\ket{\Psi_{k\Gamma_k}}_{\rm L} $ are enumerated by the index $k$ and the corresponding CM quantum numbers $\Gamma_k$ (we note that since we work in the laboratory frame, the completeness relation includes all excited states, including those with $N^{\rm CM}>0$). We introduce the Lawson term with $\lambda_{\rm CM}$ being an arbitrary (reasonably large) positive constant to help derive the t.i. Green's function, as discussed in Sec. \ref{sec:Jacobi2}. We note that for heavy targets, Eq. (\ref{GFcompl}) in laboratory coordinates can be readily used 
for evaluations of the Green's function, since the CM effects become negligible.

In what follows, we first express the overlaps in Jacobi coordinates and then provide the expression for the t.i. Green's function.

\begin{figure}[!ht]
    \centering
    {\includegraphics[width=0.5\columnwidth]{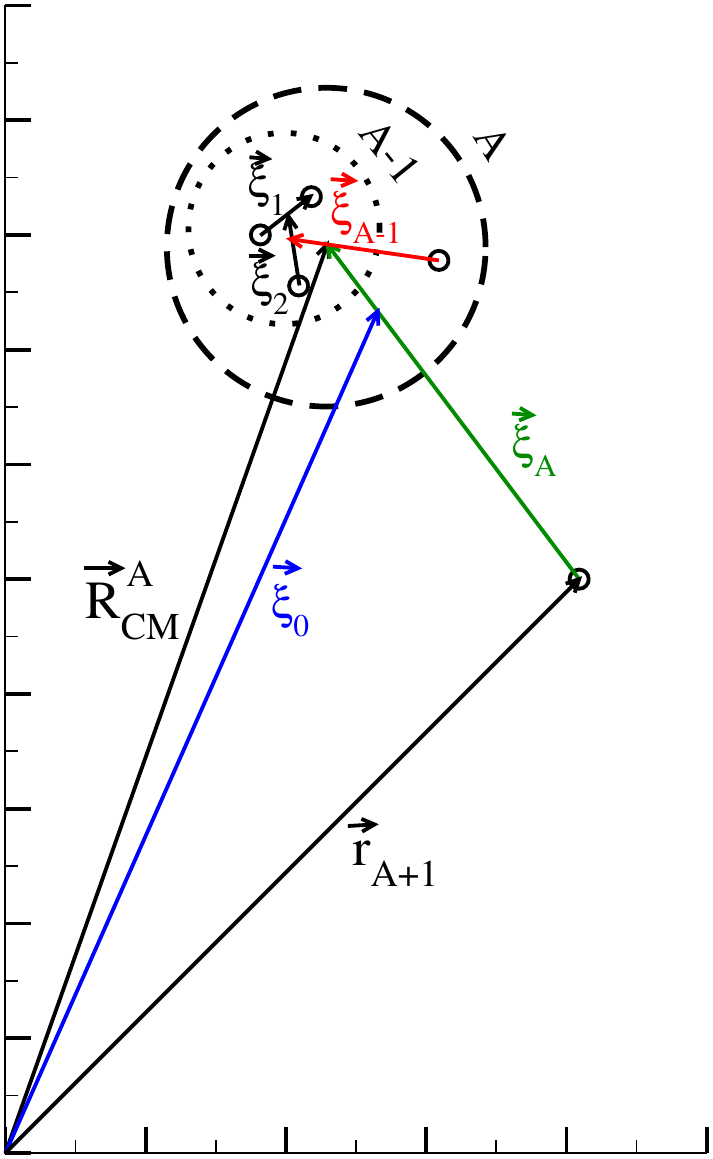}}
    \caption{A vector diagram of the Jacobi and laboratory coordinates used for the $A+1$ system (all particles), $A$ system (dashed circle), and $A-1$ system (dotted circle). Vectors are proportional to those in Eq.~(\ref{Eq.10}).}
    \label{fig:Vector_Diagram}
\end{figure}

\subsubsection{Overlaps in laboratory and Jacobi coordinates for spatial degrees of freedom}
\label{sec:Jacobi}

We utilize Jacobi coordinates (cf. Refs. \cite{Navratil:2004dp,Navratil:2004tw}), as shown in Fig. \ref{fig:Vector_Diagram}, where $\pmb{\xi}_{1,\cdots,A-1}$ are the coordinates of the nucleons in the target, $\pmb{\xi}_A$ is the relative distance between the CM of the two clusters (target and projectile), and $\pmb{\xi}_0$ is the CM coordinate of the $(A+1)$ system, with:
\begin{eqnarray}
    \pmb{\xi}_{A-1} &=& \sqrt{\frac{1}{A}} \mathbf{R}^{A-1}_{\rm CM}-\sqrt{\frac{A-1}{A}}\mathbf{r}_{A}\cr
    \pmb{\xi}_A &=& \sqrt{\frac{1}{A+1}}\mathbf{R}^{A}_{\rm CM}-\sqrt{\frac{A}{A+1}} \mathbf{r}_{A+1}\cr
    \pmb{\xi}_0 &=& \sqrt{\frac{A}{A+1}}\mathbf{R}^{A}_{\rm CM}+\sqrt{\frac{1}{A+1}} \mathbf{r}_{A+1} ,
    \label{Eq.10}
\end{eqnarray}
where ${\bf r}_{1,2,\dots,A+1}$ are laboratory coordinates of the $A+1$ nucleons, $\mathbf{R}^{A-1}_{\rm CM}=\sqrt{1/A-1}(\mathbf{r}_1+\cdots+\mathbf{r}_{A-1})$ and $\mathbf{R}^{A}_{\rm CM}=\sqrt{1/A}(\mathbf{r}_1+\cdots+\mathbf{r}_{A})$ are the laboratory coordinate of the CM of the $A-1$ and $A$ systems, respectively.

For simplicity, in this section we consider spatial dof only, with $b \equiv \{ n_b \ell_b m_b \}$, which we augment with the spin dof in Sec. \ref{sec:tiGF}. Using
$\delta(\mathbf{r}-\mathbf{r}_{A+1})= \sum_b \phi_b(\mathbf{r}) \phi^*_b(\mathbf{r}_{A+1})=$ $\sum_{n \ell m} R_{n \ell}(r) Y_{\ell m}(\hat{r})R_{n\ell}(r_{A+1})Y_{\ell m}^*(\hat{r}_{A+1}),$
the overlaps of Eq. (\ref{GFcompl}) in the laboratory frame   for general eigenfunctions of the $A$ and $A+1$ nuclei are given as:
 \begin{widetext}
\begin{eqnarray} 
    u^{A+1}_{ik(\rm L)}(\mathbf{r})&\equiv&\braketop{\Psi^{A}_{i\Gamma_i}}{a_\mathbf{r}}{\Psi^{A+1}_{k\Gamma_k}}_{\rm L}
    \cr
    &= & \sqrt{A+1} \int d \mathbf{r}_1 \dots d\mathbf{r}_{A+1}\Psi^{A}_{i\Gamma_i}(\mathbf{r}_1 ,\dots, \mathbf{r}_{A})^* 
    \Psi^{A+1}_{k\Gamma_k}(\mathbf{r}_1 ,\dots, \mathbf{r}_{A+1}) \delta(\mathbf{r}-\mathbf{r}_{A+1}) 
    =\sum_b \phi_b(\mathbf{r})  u^{A+1}_{ik,b(\rm L)}, {\rm with} \cr
    u^{A+1}_{ik,b(\rm L)} &=&\braket{(\mathcal{A} \Psi^{A}_{i\Gamma_i}  \phi_b  ) }{\Psi^{A+1}_k }_{\rm L} \nonumber \\
    &=&\sqrt{A+1} \int d \mathbf{r}_1 \dots d\mathbf{r}_{A+1}
    \Psi^{A}_{i\Gamma_i}(\mathbf{r}_1 ,\dots, \mathbf{r}_{A})^* \phi_b^*(\mathbf{r}_{A+1})  \frac{1}{\sqrt{A+1} }\sum_c \phi_c(\mathbf{r}_{A+1}) \braket{\mathbf{r}_1 ,\dots, \mathbf{r}_{A}}{a_c |\Psi^{A+1}_{k\Gamma_k}} \nonumber \\
    &=&  \braket{\Psi^{A}_{i\Gamma_i}}{a_b |\Psi^{A+1}_{k\Gamma_k}}_{\rm L}
    \equiv \braket{\Psi^{A+1}_{k\Gamma_k}}{\Phi^{+}_{i\Gamma_ib}}^*_{\rm L} .
    \label{ovlpp}
\end{eqnarray}
 \end{widetext}
(Note that the last row defines the cluster basis states for the spatial dof.) Here, $\mathcal{A}$ is the antisymmetrizer operator that ensures the antisymmetrization between the two clusters of target and projectile, which enforces the Pauli exclusion principle, and we use $\mathcal{A} \ket{\Psi^{A+1}_{k\Gamma_k}}_{\rm L} =\sqrt{A+1}\ket{\Psi^{A+1}_{k\Gamma_k}}_{\rm L} $.
This can be expressed through Jacobi coordinates (\ref{Eq.10}) (see Ref.~\cite{Navratil:2004tw}), using $\Psi^{A}_{i\Gamma_i}(\mathbf{r}_1 ,\dots, \mathbf{r}_{A}) = \Psi^{A}_i(\pmb{\xi}_1 ,\dots, \pmb{\xi}_{A-1}) \phi_{\Gamma_i}(\mathbf{R}_{\rm CM}^A)$ 
and 
$ \phi_{\Gamma_i}(\mathbf{R}_{\rm CM}^A) \phi_b(\mathbf{r}_{A+1})=\sum_{\Gamma_k \beta} \mathcal{M}^{\beta+}_{ b; \Gamma_i \Gamma_k} \phi_\beta(\pmb{\xi}_{A}) \phi_{\Gamma_k}(\pmb{\xi}_0) $ (see Appendix \ref{AppendixOverlaps}):
\begin{eqnarray}
    && u^{A+1*}_{ik,b{\rm (L)}}=
    \braket{\Psi^{A+1}_{k\Gamma_k}}{\Phi^{+}_{i\Gamma_ib}}_{\rm L} 
      =\sum_{\beta} \mathcal{M}^{\beta+}_{ b; \Gamma_i \Gamma_k} \braket{\Psi_k^{A+1}}{\Phi^{+}_{i\beta}}, \cr &&
\label{Eq.povlprel}
\end{eqnarray}
where $\mathcal{M}$ is related to the Talmi-Moshinsky bracket $\braket{n_\beta\ell_\beta N_kL_k;L}{N_iL_i n_b \ell_b;L}_{d}$ \cite{MOSHINSKY1959104,Trlifaj72,KAMUNTAVICIUS2001191}:
\begin{eqnarray}
    \mathcal{M}^{\beta+}_{ b; \Gamma_{i} \Gamma_{k}}=\sum_{LM}  \braket{n_\beta\ell_\beta N_kL_k;L}{N_iL_i n_b \ell_b;L}_{d^{+}} 
    \cr
    C_{{\ell_\beta m_\beta},{L_kM_k}}^{LM}C_{{L_iM_i},{\ell_b m_b}}^{LM}.
    \cr    \mathcal{M}^{\beta-}_{ b; \Gamma_i \Gamma_k}=\sum_{LM}  \braket{n_\beta\ell_\beta N_iL_i;L}{N_kL_k n_b \ell_b;L}_{d^-} \cr
    C_{{\ell_\beta m_\beta},{L_iM_i}}^{LM}C_{{L_kM_k},{l_b m_b}}^{LM},
    \label{TMB}
\end{eqnarray}
with $d^+=\frac{1}{A}$ for the particle states and $d^-=\frac{1}{A-1}$ for the hole states.
Similarly, for the $A-1$ system:
\begin{eqnarray}
   u^{A}_{ki(\rm L)}(\mathbf{r})&\equiv&\braketop{\Psi^{A-1}_{k \Gamma_k}}{a_\mathbf{r}}{\Psi^{A}_{i \Gamma_i}}_{\rm L} =\sum_{\beta} \phi_{\beta}(\mathbf{r})u^{A}_{ki,\beta (\rm L)}, \nonumber
\end{eqnarray}
with $u^{A}_{ki,b (\rm L)}=\braket{\Psi^{A-1}_{k \Gamma_k}}{a_b |\Psi^{A}_{i \Gamma_i}}_{\rm L}
\equiv \braket{\Psi_{k \Gamma_k}^{A-1}}{\Phi^{-}_{i\Gamma_ib}}_{\rm L}$.
The overlap can be then expressed in  Jacobi coordinates as (see Appendix \ref{AppendixOverlaps}):
\begin{eqnarray}
    u^{A*}_{ki,b (\rm L)}&=& \braket{\Psi_{k \Gamma_k}^{A-1}}{\Phi^{-}_{i\Gamma_ib}}^*_{\rm L} =\braket{ \Psi^{A}_{i\Gamma_i}}{(\mathcal{A}  \Psi^{A-1}_{k\Gamma_k } \phi_b )  }_{\rm L}   \nonumber \\
      &=& \sum_{\beta} \mathcal{M}^{\beta-}_{ b; \Gamma_i \Gamma_k}  \braket{\Psi_k^{A-1}}{\Phi^{-}_{i\beta}}^*.
      \label{hovlprel}
\end{eqnarray}

\subsubsection{Green's function for spatial degrees of freedom: Lawson procedure}
\label{sec:Jacobi2}

Using Eq.~(\ref{Eq.povlprel}) and Eq.~(\ref{hovlprel}) for the ``particle" and ``hole" overlaps, respectively, the Green's function terms of Eq.~(\ref{GFcompl}) for a given intrinsic state of the target $i$, where $i=0$ for the ground state, are expressed in Jacobi coordinates as:
\begin{eqnarray}
    &&\sum_{k \Gamma_k } \frac{ \braket{\Phi^{ \pm}_{i\Gamma_ia}}{\Psi_{k\Gamma_{k}}}_{\rm L} \braket{\Psi_{k\Gamma_{k}}}{\Phi^{\pm}_{i\Gamma_i b}}_{\rm L}  }{E - (\varepsilon_k^\pm \pm \lambda_{\rm CM} N_{k}^{\rm CM} ) \pm  \iu\epsilon} \nonumber \\ 
    &=&\sum_{k \alpha \beta} \left[ \sum_{\Gamma_{k}} \frac{ \mathcal{M}^{\alpha \pm}_{ a; \Gamma_i \Gamma_k}\mathcal{M}^{\beta \pm}_{ b; \Gamma_i\Gamma_k}
    }{E - (\varepsilon_k^\pm \pm \lambda_{\rm CM} N_{k}^{\rm CM} ) \pm  \iu\epsilon} \right]
    \braket{\Phi^{\pm}_{i \alpha} }{\Psi_k} \braket{\Psi_k}{\Phi^{\pm}_{i \beta}},  \nonumber \\
    \label{GFnboth}
\end{eqnarray}
where $\Gamma_i$ and $\Gamma_k$ denote the center-of-mass labels for the target and $A\pm 1$ states, respectively, and $i$ and $k$ denote the corresponding intrinsic-state labels.
Taking the limit of large (but finite) $\lambda_{\rm CM}$ on both sides of Eq. (\ref{GFnboth}):
\begin{eqnarray}
    &&\sum_{k } \frac{ \braket{\Phi^{\pm}_{i\Gamma_ia}}{\Psi_{k,\Gamma_k=0}}_{\rm L} \braket{\Psi_{k,\Gamma_k=0}}{\Phi^{\pm}_{i\Gamma_ib}}_{\rm L}  }{E - \varepsilon_k^\pm \pm  \iu\epsilon}
     \nonumber \\ 
    &=&\sum_{k \alpha \beta} \left[  \frac{ \mathcal{M}^{\alpha\pm}_{ a; \Gamma_i 0}\mathcal{M}^{\beta\pm}_{ b; \Gamma_i 0}
    }{E - \varepsilon_k^\pm \pm  \iu\epsilon} \right]  \braket{\Phi^{\pm}_{i \alpha}}{\Psi_k} \braket{\Psi_k}{\Phi^{\pm}_{i\beta}},\nonumber \\ 
    &&({\rm large}\, \lambda_{\rm CM}, |E|<\lambda_{\rm CM}),
    \label{GFnboth2}
\end{eqnarray}
where the only terms that are nonzero are those with $\Gamma_k=0$ (that is, $N_k^{\rm CM}=0$, $L_k^{\rm CM}=0$, and $M_{k}^{\rm CM}=0$). 
The condition $|E|<\lambda_{\rm CM}$ ensures that we exclude the poles of the CM spurious states that affect the imaginary part, as discussed below. This condition on $E$ is always valid for the low-energy regime of applicability of this approach (usually, $\lambda_{\rm CM}\sim 50-100$ MeV).

Equation (\ref{GFnboth2}) is true for \emph{any} $\Gamma_i$ of the target, and we can choose, without loss of generality (w.l.g.), $\Gamma_i=0$ implying $a=\alpha$ and $b=\beta$:
\begin{eqnarray}
    &&\sum_{k } \frac{ \braket{\Phi^{\pm}_{i,\Gamma_i=0,a}}{\Psi_{k,\Gamma_k=0}}_{\rm L} \braket{\Psi_{k,\Gamma_k=0}}{\Phi^{\pm}_{i,\Gamma_i=0,b}}_{\rm L}  }{E - \varepsilon_k^\pm \pm  \iu\epsilon}\nonumber \\ 
    &=&
    \mathcal{M}^{a\pm}_{ a; 0 0}\mathcal{M}^{b\pm}_{b; 00}
   \left[ \sum_k \frac{ \braket{\Phi^{\pm}_{ia}}{\Psi_k} \braket{\Psi_k}{\Phi^{\pm}_{ib}}
    }{E - \varepsilon_k^\pm  \pm  \iu\epsilon} \right],\nonumber \\ 
    &&({\rm large}\, \lambda_{\rm CM}, |E|<\lambda_{\rm CM}).
    \label{tiGFa}
\end{eqnarray}
The quantity in the brackets is exactly the t.i. particle or hole term of the Green's function. We emphasize  that Eq. (\ref{tiGFa}) does not restrict the target state to no CM motion in the reaction dynamics but rather is a mathematical equality that uses only wavefunctions free of CM excitations to exactly connect the SA-NCSM calculations  to the t.i. counterparts that can be, in general, calculated in Jacobi coordinates. This is possible because the entire information for the intrinsic function is contained in the case of $\Gamma=0$. 

To use the completeness relationship in Eq. (\ref{tiGFa}) and express the Green's function in an operator form, we 
use the projection of the  cluster basis state  $\ket{\Phi^{\pm}_{i\Gamma_ib}}_{\rm L} $ to its component that is free of CM excitations, $\ket{\Phi^{0\pm}_{i\Gamma_i b}}_{\rm L} $, which we denote by the ``0" superscript [see Eq. (\ref{projCM})]. We can now 
use that $\braket{\Psi_{k, \Gamma_k=0}}{\Phi^{\pm}_{i \Gamma_i b}}_{\rm L}=\braket{\Psi_{k\Gamma_{k}}}{\Phi^{0\pm}_{i \Gamma_i b}}_{\rm L}$.
This is important since the completeness relation includes all intrinsic states $k$ in the intrinsic frame but requires both $k$ and $\Gamma_k$ in the laboratory frame. Starting from the r.h.s. of Eq. (\ref{tiGFa}), the transitionally invariant Green's function terms  are thus given as (with $i=0$ for the ground state):
\begin{eqnarray}
    && \braketop{\Phi^{\pm}_{i\alpha}}{
    \frac{1}{E - (\hat H-E_i^A -  \iu\epsilon)(\hat N-A)}}{\Phi^{\pm}_{i\beta}}
    \nonumber \\ 
    &=&  \sum_k \frac{ \braket{\Phi^{\pm}_{i\alpha}}{\Psi_k} \braket{\Psi_k}{\Phi^{\pm}_{i\beta}}
    }{E - \varepsilon_k^\pm  \pm  \iu\epsilon} \nonumber \\
    &=& \frac{1}{ \mathcal{M}^{\alpha\pm}_{ \alpha; 0 0}\mathcal{M}^{\beta\pm}_{\beta; 00} }\sum_{k \Gamma_k} \frac{ \braket{\Phi^{0\pm}_{i0\alpha}}{\Psi_{k\Gamma_k}}_{\rm L} \braket{\Psi_{k\Gamma_k}}{\Phi^{0\pm}_{i0\beta}}_{\rm L}  }{E - \varepsilon_k^\pm \pm  \iu\epsilon}\nonumber \\ 
    &=&\frac{1}{ \mathcal{M}^{\alpha\pm}_{ \alpha; 0 0}\mathcal{M}^{\beta\pm}_{\beta; 00} } \cr
    &\times &
   \braketop{\Phi^{0\pm}_{i0\alpha}}{
    \frac{1}{E - (\hat H +\lambda_{\rm CM} \hat N_{\rm CM} - E_i^A -  \iu\epsilon)(\hat N-A)}}{\Phi^{0\pm}_{i0\beta}}_{\rm L}\nonumber \\ 
    &&({\rm large}\, \lambda_{\rm CM}, |E|<\lambda_{\rm CM}).
    \label{tiGF}
\end{eqnarray}
Hence, in general, the t.i. Green's function is calculated using the last part of Eq. (\ref{tiGF}) that can employ any many-body method with laboratory coordinates\footnote{
We note that, in general, the many-body method employed needs to ensure that the center-of-mass wavefunction is factored out exactly, as in the case of NCSM and SA-NCSM, or \emph{near} exactly but with an error estimate for the CM contamination.
} and the Lawson procedure. Most importantly, for these calculations we use cluster basis states $\ket{\Phi^{0\pm}_{i\Gamma_i \alpha}}_{\rm L} $ that have no CM excitations (see Sec. \ref{section:Ortho_CM_Removal} for details on removing the CM excitations from these states). 

Furthermore, this implies that we need simple Talmi-Moshinsky brackets, since $\Gamma_i=\Gamma_k=0$ (see, e.g., \cite{Navratil:2004tw}): 
\begin{equation}
        \mathcal{M}_{\alpha;00}^{\alpha\pm }=\braket{n_\alpha \ell_\alpha 00;\ell_\alpha}{00 n_\alpha \ell_\alpha;\ell_\alpha}_{d^\pm} \
    = (-1)^{\ell_\alpha} \left( \frac{1}{1+d^\pm} \right)^{n_\alpha/2},
\end{equation}
with $\mathcal{M}_{\alpha;00}^{\alpha + }=(-1)^{\ell_\alpha} \left( \frac{A}{A+1} \right)^{n_\alpha/2}$ for the particle case and  $\mathcal{M}_{\alpha;00}^{\alpha - }=(-1)^{\ell_\alpha} \left( \frac{A-1}{A} \right)^{n_\alpha/2}$ for the hole case.

\subsubsection{Translationally invariant Green's function for spatial-spin degrees of freedom}
\label{sec:tiGF}
Importantly, for $\Gamma_i=\Gamma_k=0$, the generalization to spatial-spin degrees of freedom with $\alpha=n_\alpha(\ell_\alpha \half)j_\alpha$ is straightforward for Eq.~(\ref{tiGF}), since the coupling to the CM wavefunction is trivial (omitting $i$ and $\Gamma_i=0$ from the notations):
\begin{eqnarray}
    &&
    \braketop{\Phi^{J\pm}_{\alpha}}{\hat{G}(E,\epsilon)}{\Phi^{J\pm}_{\beta}}
    = \frac{
    \braketop{\Phi^{J0\pm}_{\alpha}}{\hat{G}(E,\epsilon)}{\Phi^{J0\pm}_{\beta}}_{\rm L}
    }{ \mathcal{M}^{\alpha\pm}_{ \alpha; 0 0}\mathcal{M}^{\beta\pm}_{\beta; 00} }
   \nonumber \\ 
    &&({\rm large}\, \lambda_{\rm CM}, |E|<\lambda_{\rm CM}),
    \label{tiGFlsj}
\end{eqnarray}
where the operator ${\hat{G}(E,\epsilon)}$ is defined in Eq. (\ref{Gop})\footnote{
Similarly to Eq.~(\ref{tiGF}), the Lawson term is included on the r.h.s. of Eq. (\ref{tiGFlsj}) and is used for all present calculations in the laboratory frame, although it is omitted from the notations given the trivial contribution, with ${\lambda_{\rm CM} \hat N_{\rm CM} }\ket{\Phi^{J0\pm}_{\beta}}=0$.
}. 
We emphasize the use of $\ket{\Phi^{J0\pm}_{\beta}}_{\rm L}$ cluster basis states that have no CM excitations. While this provides the most general way to calculate the t.i. Green's function, some of the conditions can be relaxed depending on the choice of the frame, as discussed next.

In what follows, we will distinguish between the $E>\varepsilon_F^+$ regime from the $E<\varepsilon_F^-$ regime, where the $\varepsilon_F$ energies define the single-nucleon thresholds, $\varepsilon_F^+ \equiv E_0^{A+1}-E_0^A$ and  $\varepsilon_F^- \equiv E_0^A- E_0^{A-1}$. The reason is that the calculations are performed in the CM reference frame of the $A+1$ system for $E>\varepsilon_F^+$ and of the $A-1$ system for $E<\varepsilon_F^-$.

\vspace{12pt}
{\textit{$E \ge \varepsilon_F^+$ Regime:}} 
For these energies (relevant to particle-target reactions processes), the problem is solved in the $A+1$ CM system with reduced mass $\mu_p=\frac{A}{A+1}m_N$ (and HO characteristic length $b_p=\sqrt{\frac{\hbar}{\mu_p \Omega }}$), 
where $m_N$ is the nucleon mass. Hence, by construction, $\Gamma_k=0$ for $\ket{\Psi^{A+1}_{k \Gamma_k}}_{\rm L} $, and one can use Eq.~(\ref{tiGFlsj}) for the ``particle" term without the need for the Lawson technique or restriction on the energy, simply based on the transformation of the overlaps to Jacobi coordinates [see Eq. (\ref{projp})]:
\begin{eqnarray}
    &&
\sum_k \frac{ \braket{\Phi^{J+}_{\alpha}}{\Psi_k} \braket{\Psi_k}{\Phi^{J+}_{\beta}}
    }{E - \varepsilon_k^+  +  \iu\epsilon} \nonumber \\
   &=&(-)^{\ell_\alpha-\ell_{\beta}}\left(\frac{A+1}{A}\right)^{\frac{n_\alpha+n_{\beta}}{2}} 
    \sum_{k } \frac{ \braket{\Phi^{J0+}_{\alpha}}{\Psi_{k0}}_{\rm L} \braket{\Psi_{k0}}{\Phi^{J0+}_{\beta}}_{\rm L}  }{E - \varepsilon_k^+ +  \iu\epsilon} \nonumber \\
    &=&\left(\frac{A+1}{A}\right)^{\frac{n_\alpha+n_{\beta}}{2}} 
    \braketop{\Phi^{J0+}_{\alpha}}{\hat{G}(E,\epsilon)}{\Phi^{J0+}_{\beta}}_{\rm L},
    \label{tiGFp}
\end{eqnarray}
where we use the parity conservation $\pi_{J_0}(-1)^{\ell_\alpha}=\pi_J=\pi_{J_0}(-1)^{\ell_{\beta}}$ and use again that $\braket{\Psi_{k0}}{\Phi^{J0+}_{\beta}}_{\rm L}=\braket{\Psi_{k \Gamma_{k}}}{\Phi^{J0+}_{\beta}}_{\rm L} $ to ensure the completeness relation in the laboratory frame. We note that $\ket{\Phi^{J0+}_{\beta}}_{\rm L}$ is an $(A+1)$-body state and, by construction, has no CM excitations with respect to the CM of the $A+1$ system, which in this case coincides with the frame of choice.

 Using Eq. (\ref{tiGFp}) and the practical identity $\lim_{\epsilon \rightarrow 0} \frac{1}{E\pm \iu\epsilon}={\rm p.v.} \frac{1}{E} \mp i \pi \delta(E)$, where  ${\rm p.v.}$ denotes the principal value (and now we explicitly include $\lim$ for clarity), we obtain the t.i. Green's function used in the present calculations for projectile-target reaction processes:
\begin{eqnarray}
    &&
G^{J+}_{\alpha \beta;E \ge \varepsilon_F^+}\equiv\lim_{\epsilon \rightarrow 0} \sum_k \frac{ \braket{\Phi^{J+}_{\alpha}}{\Psi_k} \braket{\Psi_k}{\Phi^{J+}_{\beta}}
    }{E - \varepsilon_k^+  +  \iu\epsilon} \nonumber \\
    &=&\lim_{\epsilon \rightarrow 0} \left(\frac{A+1}{A}\right)^{\frac{n_\alpha+n_{\beta}}{2}}  
    \braketop{\Phi^{J0+}_{\alpha}}{\hat{G}(E,\epsilon)}{\Phi^{J0+}_{\beta}}_{\rm L}
 \label{tiGFp_pT} \\
   &=&\left(\frac{A+1}{A}\right)^{\frac{n_\alpha+n_{\beta}}{2}}  
   \left[
     {\rm p.v.} \sum_{k \Gamma_{k}} \frac{ \braket{\Phi^{J0+}_{\alpha}}{\Psi_{k\Gamma_{k}}}_{\rm L} \braket{\Psi_{k\Gamma_{k}}}{\Phi^{J0+}_{\beta}}_{\rm L}  }{E - \varepsilon_k^+} \right. \nonumber \\
     &&\left. -i\pi \sum_{k\Gamma_{k}} \braket{\Phi^{J0+}_{\alpha}}{\Psi_{k\Gamma_{k}}}_{\rm L} \braket{\Psi_{k\Gamma_{k}}}{\Phi^{J0+}_{\beta}}_{\rm L}\delta(E - \varepsilon_k^+)
     \right],
    \label{tiGFp_pv}
\end{eqnarray}
where the principal value, 
\scalebox{.88}[1.0]{
${\rm p.v.}  \braketop{\Phi^{J0+}_{\alpha}}{\hat{G}(E,\epsilon=0)}{\Phi^{J0+}_{\beta}}_{\rm L}$,
}
can be straightforwardly calculated for $E$ outside $[\varepsilon_k^+-\Delta \varepsilon,\varepsilon_k^++\Delta \varepsilon]$ for small (and finite) $\Delta \varepsilon$ energy interval.
The last term in (\ref{tiGFp_pv}) when integrated  yields the t.i. ``particle" norm \cite{DickhoffBook}, 
\begin{eqnarray}
    \mathcal{N}^p_{\alpha \beta}&=&-\frac{1}{\pi}\int_{\varepsilon_F^+}^{\infty} {\rm Im}(G^{J+}_{\alpha \beta;E>\varepsilon_F^+}) dE \cr
    &=&\left(\frac{A+1}{A}\right)^{\frac{n_\alpha+n_{\beta}}{2}}
    \braket{\Phi^{J0+}_{\alpha}}{\Phi^{J0+}_{\beta}}_{\rm L}.
    \label{Eq.Particle_Norm}
\end{eqnarray}
This coincides with the t.i. norm of the resonating group method (RGM) approach \cite{QuaglioniN09}, but here it is calculated in a different way through the CM-excitations-free cluster basis states  (Appendix \ref{appendix:B}).

However, for this regime, the $A-1$ system is not in the lowest CM state. Fortunately, for the ``hole" part one can use Eq.~(\ref{tiGFlsj}) [based on Eq. (\ref{tiGF})], since the Lawson technique enables exact solutions, that, importantly, require  only the calculation of $\ket{\Phi^{J0-}_{\beta}}_{\rm L} $. We note that these cluster basis states are free from CM spurious motion with respect to the CM of the $A-1$ system, which do not coincide with the frame of choice, but become useful in the Lawson procedure (\ref{tiGF}):
\begin{eqnarray}
    &&
G^{J-}_{\beta \alpha ;E \ge \varepsilon_F^+}\equiv\lim_{\epsilon \rightarrow 0} \sum_k \frac{ \braket{\Phi^{J-}_{\beta}}{\Psi_k} \braket{\Psi_k}{\Phi^{J-}_{\alpha}}
    }{E - \varepsilon_k^-  +  \iu\epsilon} \nonumber \\
  &&=  \lim_{\epsilon \rightarrow 0} \left(\frac{A}{A-1}\right)^{\frac{n_\alpha+n_{\beta}}{2}}
  \braketop{\Phi^{J0-}_{\beta}}{\hat{G}(E,\epsilon)}{\Phi^{J0-}_{\alpha}}_{\rm L} \nonumber \\
   &&=\left(\frac{A}{A-1}\right)^{\frac{n_\alpha+n_{\beta}}{2}} \left[
     {\rm p.v.}  \braketop{\Phi^{J0-}_{\beta}}{\hat{G}(E,\epsilon=0)}{\Phi^{J0-}_{\alpha}}_{\rm L} \right. \nonumber \\
     &&\left. +i\pi \sum_{k\Gamma_{k}} \braket{\Phi^{J0-}_{\beta}}{\Psi_{k\Gamma_{k}}}_{\rm L} \braket{\Psi_{k\Gamma_{k}}}{\Phi^{J0-}_{\alpha}}_{\rm L}\delta(E - \varepsilon_k^-)
     \right]\nonumber \\ 
    &&({\rm large}\, \lambda_{\rm CM}),
    \label{tiGFh}
\end{eqnarray}
which holds for any $E$ since $\varepsilon_F^+>-\lambda_{\rm CM}$ is always the case\footnote{
In general, for any $J_0$ and for $E \ge \varepsilon_F^+$:
\begin{eqnarray}
G^J_{\alpha\beta}  
&=&G^{J+}_{\alpha\beta}+ (-1)^{2J_0+1}\sum_{J'}\Pi^2_{J'}
 \sixj{j_\alpha}{J_0}{J'}{j_\beta}{J_0}{J}G^{J'-}_{\beta\alpha}, \nonumber
 \label{GFconfig2nonzeroJ0}
\end{eqnarray}
which for $J_0=0$ coincides with Eqs. (\ref{GFconfig}) \& (\ref{GFconfig2}). This relation holds for the norm when the identity operator is used, with $\delta_{\alpha \beta}$ on the l.h.s. (cf. Ref. \cite{Birse:1981ghl}).
}. 
 As discussed in Sec. \ref{sec:Jacobi2}, the intrinsic operator $\hat{G}(E,\epsilon)$ acts here only on the intrinsic structure of the $A-1$ system, and hence its t.i. matrix elements can be obtained by pushing the states with CM excitations to high energies that no longer contribute to the matrix elements (we emphasize that the Lawson procedure is suitable to the Green's function operator due to the inverse dependence on the t.i. $A-1$ Hamiltonian, but cannot be applied to any operator in general).

In this energy regime, the t.i. ``hole" norm cannot be calculated through the imaginary part since $E \ge \varepsilon_F^+$, but is readily derived through the particle norm, $\mathcal{N}^h_{\beta \alpha}=\delta_{\alpha \beta}-\mathcal{N}^p_{\alpha \beta}$ [cf. Eq. (\ref{Eq:Norm})].

\textit{$E \le \varepsilon_F^-$ Regime:} For these energies (relevant to knock-out reactions), the problem is solved in the $A-1$ CM system with reduced mass $\mu_h=\frac{A-1}{A}m_N$ (and HO characteristic length $b_h=\sqrt{\frac{\hbar}{\mu_h \Omega }}$). Hence, by construction, $\Gamma_k=0$ for $\ket{\Psi^{A-1}_{k \Gamma_k}}_{\rm L} $, which is now considered with respect to the $A-1$ CM system. In this case, the ``hole" part can be straightforwardly calculated, while the CM motion of the $A+1$ system requires the use of the Lawson procedure. That is, one can use Eqs.~(\ref{tiGFp_pT}) and (\ref{tiGFp_pv}), but for the ``hole" term in the Green's function without the need for the Lawson technique or restriction on the energy:
\begin{eqnarray}
    &&
G^{J-}_{\beta \alpha ;E \le \varepsilon_F^-}\equiv\lim_{\epsilon \rightarrow 0} \sum_k \frac{ \braket{\Phi^{J-}_{\beta}}{\Psi_k} \braket{\Psi_k}{\Phi^{J-}_{\alpha}}
    }{E - \varepsilon_k^-  -  \iu\epsilon} \nonumber \\
    &&=\lim_{\epsilon \rightarrow 0} 
    \left(\frac{A}{A-1}\right)^{\frac{n_\alpha+n_{\beta}}{2}} 
    \braketop{\Phi^{J0-}_{\beta}}{\hat{G}(E,\epsilon)}{\Phi^{J0-}_{\alpha}}_{\rm L}.
    \label{tiGFh_negE}
\end{eqnarray}
The imaginary part of this when integrated yields the t.i. ``hole" norm (with respect to the CM of the $A-1$ system, the frame of choice in this case):
\begin{eqnarray}
    \mathcal{N}^h_{\beta \alpha }&=&\frac{1}{\pi}\int_{-\infty}^{\varepsilon_F^-} {\rm Im}(G^{J-}_{\beta \alpha;E<\varepsilon_F^-}) dE \cr
    &=& \left(\frac{A}{A-1}\right)^{\frac{n_\alpha+n_{\beta}}{2}} 
    \braket{\Phi^{J0-}_{\beta}}{\Phi^{J0-}_{\alpha}}_{\rm L}.
    \label{Eq.Hole_Norm}
\end{eqnarray}
Similarly, the Lawson technique is used for the particle side by employing Eq.~(\ref{tiGFlsj}), 
\begin{eqnarray}
    &&
G^{J+}_{\alpha \beta;E \le \varepsilon_F^-}\equiv\lim_{\epsilon \rightarrow 0} \sum_k \frac{ \braket{\Phi^{J+}_{\alpha}}{\Psi_k} \braket{\Psi_k}{\Phi^{J+}_{\beta}}
    }{E - \varepsilon_k^+  -  \iu\epsilon} \nonumber \\
  &&=  \lim_{\epsilon \rightarrow 0} 
  \left(\frac{A+1}{A}\right)^{\frac{n_\alpha+n_{\beta}}{2}}  \braketop{\Phi^{J0+}_{\alpha}}{\hat{G}(E,\epsilon)}{\Phi^{J0+}_{\beta}}_{\rm L} 
  \nonumber \\ 
    &&({\rm large}\, \lambda_{\rm CM}),
    \label{tiGFp_lim}
\end{eqnarray}
which holds for any $E$ since $\varepsilon_F^-<\lambda_{\rm CM}$ is always the case.
Similarly to above, in this energy regime, the t.i. ``particle" norm cannot be calculated through the imaginary part since $E\le \varepsilon_F^-$, but is readily derived through the hole norm, $\mathcal{N}^p_{\alpha \beta}=\delta_{\alpha \beta}-\mathcal{N}^h_{\beta \alpha}$ [cf. Eq. (\ref{Eq:Norm})].
 
Clearly, in the case of heavy targets ($A \gg 1$), the $A$-dependent CM factors become unity (equivalent to a target with no recoil, for which the laboratory and CM frames coincide) and $G^\pm$ for $E \ge \varepsilon_F^+$ coincide with those for $E \le \varepsilon_F^-$, reproducing earlier formulae that neglect the CM effects \cite{DickhoffBook,RotureauDHNP17}.

\subsection{Orthonormalization and removing CM spuriosity in cluster basis states}
\label{section:Ortho_CM_Removal}

As emphasized in the preceding section, the translationally invariant Green's function is calculated for cluster basis states $\ket{\Phi^{J0\pm}_{a}}_{\rm L}$ defined in Eq.~(\ref{Eq:pivots}) that, in addition, have no spurious CM excitations with respect to the CM of the $A\pm 1$ system. 

To achieve this, we perform the following steps: {\bf 1.} We solve the Schr\"odinger equation for the $J_0$ ground-state of the target (omitting the $J_0$ notation), $(\hat H + \lambda_{\rm CM}\hat{N}_{\rm CM}) \ket{\Psi_0^A}_{\rm L} =E_0^A \ket{\Psi_0^A}_{\rm L}  $, which yields $\ket{\Psi_0^A}_{\rm L} $ with no spurious CM excitations (ensuring $\Gamma_0=0$ discussed above). {\bf 2.} We generate the cluster basis states according to Eq.~(\ref{Eq:pivots}). 
{\bf 3.} We ensure an orthonormal basis for which 
$\mathcal{N}_{ab(L)}^p+\mathcal{N}_{ba(L)}^h=\delta_{ab}$ according to Eq. (\ref{Eq:Norm}), which for $J_0=0$ implies $\braket{\Phi^{J+}_{a}}{\Phi^{J+}_{b}}_{\rm L}+\braket{\Phi^{J-}_{b}}{\Phi^{J-}_{a}}_{\rm L}=\delta_{ab}$ or equally $a_a a^\dagger_b+a^\dagger_b a_a=\delta_{ab}$.
For each $\ell_a=\ell_b$ and $j_a=j_b$, this equality  holds approximately, especially for higher $n_a$ ($n_b$) shells, due to the use of $N_{\rm max}$ in the no-core shell-model-type calculations. To address this, we orthonormalize the basis by calculating the total norm 
for each $J$, $\mathcal{N}_{ab(L)}=\mathcal{N}_{ab(L)}^p+\mathcal{N}_{ba(L)}^h$
and $\ket{\bar \Phi^{J\pm}_{a}}_{\rm L} =\sum_{n_b}\mathcal{N}^{-1/2}_{ab(L)} \ket{\Phi^{J\pm}_{b}}_{\rm L} $, with $\mathcal{N}^{-1/2}_{(L)}$ calculated through the $\mathcal{N}_{(L)}$ eigenvalues and eigenvectors. Therefore, calculations are performed in the $\ket{\bar \Phi^{J}}$ orthonormal basis in the laboratory frame and, hence, in the intrinsic frame
(this further implies that the inverse of the Green's function can be calculated through $ \mathbf{G}^{-1} \mathbf{G}=\mathds{1}$). Cluster basis states used henceforth are \textit{orthonormalized} and we will omit the bar symbol from their notations.  
{\bf 4.} We remove CM excitations in the orthonormalized basis $\ket{\Phi^{J0\pm}_{a}}_{\rm L}$ through a projection method utilized in an earlier study \cite{BakerLBND20}. This is achieved by applying a projection operator $\hat{\mathcal{P}}_0$, such that $\ket{\Phi^{J0\pm}_{a}}_{\rm L}= \hat{\mathcal{P}}_0 \ket{\Phi^{J\pm}_{a}}_{\rm L}$, where
\begin{eqnarray}
    \hat{\mathcal{P}}_0 = \prod_{N^{\rm CM}=1}^{N_{\rm max}} \left( \mathds{1} - \frac{\hat{N}_{\rm CM}}{N^{\rm CM}} \right).
    \label{projCM}
\end{eqnarray}
 The $\ket{\Phi^{J0\pm}_{a}}_{\rm L}$ cluster basis states are then used to calculate $\braketop{\Phi^{J0\pm}_{a}}{\hat{G}(E,\epsilon)}{\Phi^{J0\pm }_{b}}_{\rm L} $ matrix elements for the t.i. Green's function (Sec. \ref{sec:tiGF}) through the Lanczos method, discussed next. 

\subsection{Lanczos method for Green's function}
\label{section:Lanczos}

In the Lehmann representation, one can calculate $\ket{\Psi_{\Gamma_k k}^{A\pm 1}}_{\rm L} $ from
$(\hat H + \lambda_{\rm CM}\hat{N}_{\rm CM}) \ket{\Psi_{\Gamma_k k}^{A\pm 1}}_{\rm L} =E_k^{A\pm 1} \ket{\Psi_{\Gamma_k k}^{A\pm 1}}_{\rm L} $ and use these to compute the overlaps in
Eq.~(\ref{GFcompl}).  
However, as noted in Ref. \cite{RotureauDHNP17},  what one finds in practice is that this method is slow at converging with respect to the number of Lanczos iterations due to the number of eigenvectors needed. 

Instead, the Green's function matrix elements are calculated through a Lanczos method based on a continued fraction evaluation similar to that performed in Lorentz-Integral-Transformation (LIT) calculations \cite{Marchisio2003}. Specifically, $G^{J+}_{\alpha\beta ({\rm L})}(z) \equiv \braketop{\Phi^{J0+}_{ \alpha}}{\hat{G}(E,\epsilon)}{\Phi^{J0+ }_{ \beta}}_{\rm L} $  in Eqs.~(\ref{tiGFp_pT}) and (\ref{tiGFp_lim}) is computed as:
\begin{eqnarray}
    G^{J+}_{\alpha\beta({\rm L})}(z) = \sum_{k=0}^{N_{\rm iter}} \left\langle \Phi^{J0+}_{\alpha} | q_k \right\rangle \left\langle q_k \left| \frac{1}{z-\hat H} \right| \Phi^{J0+}_{\beta} \right\rangle ,
    \label{GLanczos1}
\end{eqnarray}
where $z=E+E_0^A+\iu\epsilon$, $\ket{q_k}$ are the Lanczos vectors, where $k$ goes from 0 to the number of Lanczos iterations $N_{\rm iter}$, and the completeness of the  Lanczos vectors $\sum_k \left| q_k \right\rangle \left\langle q_k \right| \simeq 1$ is inserted (all calculations in this section are performed for wavefunctions in the laboratory frame, thereby omitting ${\rm L}$ from the notations; also, the use of a Lawson term that augments $\hat H$ should be understood). The approximation in the completeness relation depends on the number of Lanczos iterations and is practically negligible for a sufficiently large number, as discussed at the end of this subsection. Specifically, in this study, we use $N_{\rm iter}= 2000$  or the complete basis size for a given $N_{\rm max}$, if smaller.

The Lanczos algorithm starts with a so-called pivot vector $\ket{q_0}$,
which for the Green's function calculations corresponds to  $\left| \Phi^+ \right\rangle$ for $G^+$ (or $\left| \Phi^- \right\rangle$ for $G^-$). Specifically, the algorithm uses \textit{normalized} pivots as input, that is, 
\begin{equation}
\left| q_0^{\beta} \right\rangle \equiv \frac{\left| \Phi^{J0+}_{\beta} \right\rangle}{\sqrt{\lla \Phi^{J0+}_{\beta} | \Phi^{J0+}_{\beta} \rra}}.     
\end{equation}
Hence, Eq. (\ref{GLanczos1}) can be equivalently written as
\begin{eqnarray}
    G^{J+}_{\alpha\beta({\rm L})}(z) &=&  \sqrt{\lla \Phi^{J0+}_{\beta} | \Phi^{J0+}_{\beta} \rra} \cr
    & & \times \sum_k \left\langle \Phi^{J0+}_{\alpha} | q_k \right\rangle \left\langle q_k \left| \frac{1}{z-\hat H} \right| q_0^\beta
    \right\rangle.
    \label{eq:40}
\end{eqnarray}
To calculate the matrix elements,
\begin{eqnarray}
     x_{k0}^\beta \equiv \left\langle q_k \left| \frac{1}{z-\hat H} \right| q_0^\beta\right\rangle = x_{0k}^\beta~,
\end{eqnarray}
the continued fraction evaluation, based on Cramer's rule \cite{Marchisio2003}, is used:
\begin{eqnarray}
    x_{0k} &=& \frac{1}{(z-a_k)\lambda_{0 k-1} - b_{k} \lambda_{0 k-2} + \lambda_{0 k-1} g_{k+1}} \cr \cr \cr
	\lambda_{0k} &=& \frac{(z-a_k)\lambda_{0 k-1} - b_k \lambda_{0 k-2}}{b_{k+1}} \cr \cr \cr
	g_k &=& \frac{-b_k^2}{(z-a_k)-\frac{b_{k+1}^2}{(z-a_{k+1})-\frac{b_{k+2}^2}{\cdots}}},
 \label{contfrac}
\end{eqnarray}
where $a_k$ and $b_k$, $k=0,1,\dots, N_{\rm iter}$, are the diagonal and off diagonal matrix elements of the tridiagonal Lanczos matrix, respectively, commonly referred to as Lanczos coefficients. At a given Lanczos iteration $k$, the continued fraction in $g_k$ depends on the Lanczos coefficients $a_k,b_k,a_{k+1},b_{k+1},\dots a_{N_{\rm iter}},$ and $b_{N_{\rm iter}}$; $x_{0k}$ is calculated recursively, with a base case  $\lambda_{-1}=1$ and $\lambda_{-2}=0$. Fortunately, one does not need to calculate all $N_{\rm iter}$ matrix elements of  $x_{0k}$ but much fewer (typically, about 200 or less) till the sum $\sum_k \left\langle \Phi^{J0+}_{\alpha} | q_k \right\rangle x_{k0}^{\beta}$ in $G^{J+}_{\alpha\beta({\rm L})}(z) $  becomes converged within a precision level ($10^{-16}$ used in the present calculations). 
Furthermore, for the case where $\alpha=\beta$, $G^{J+}_{\alpha\alpha({\rm L})}(z)$, only $x_{00}$ is needed,
\begin{eqnarray}
    x_{00} =  \frac{1}{(z-a_0) -  \frac{b_1^2}{(z-a_1) -  \frac{b_2^2}{(z-a_2) - \frac{b_3^2}{\cdots}} } },
\end{eqnarray}
since the Lanczos vectors are orthonormal $\braket{q_\alpha}{q_\beta}=\delta_{\alpha\beta}$.

The Green's function $G^{J-}_{\beta \alpha({\rm L})}(z)  \equiv \braketop{\Phi^{J0-}_{ \beta}}{\hat{G}(E,\epsilon)}{\Phi^{J0- }_{ \alpha}}_{\rm L} $ in Eqs.~(\ref{tiGFh}) and (\ref{tiGFh_negE}) can be evaluated in exactly the same manner, using the recursion relation:
\begin{eqnarray}
    x^-_{0k} &=& (-1)^k\frac{1}{(z+a_k)\lambda^-_{0 k-1} - b_{k} \lambda^-_{0 k-2} + \lambda^-_{0 k-1} g^-_{k+1}} \cr \cr \cr
	\lambda^-_{0i} &=& \frac{(z+a_k)\lambda^-_{0 k-1} - b_k \lambda^-_{0 k-2}}{b_{k+1}} \cr \cr \cr
	g^-_k &=& \frac{-b_i^2}{(z+a_k)-\frac{b_{k+1}^2}{(z+a_{k+1})-\frac{b_{k+2}^2}{\cdots}}},
\end{eqnarray}
where $z=E-E_0^A - \iu\epsilon$.
Alternatively, one can use Eqs. (\ref{eq:40}) and (\ref{contfrac}) but with $z=-E+E^A_0+\iu\epsilon$ to compute $-G^{J-}_{\beta \alpha({\rm L})}(z)$.

Finally, the Lanczos vector completeness relation can be tested using $\delta_{m0}=\sum_k \left\langle q_m \left| z-\hat H \right| q_k \right\rangle x_{k0} $, where the completeness of the  Lanczos vectors $\sum_k \left| q_k \right\rangle \left\langle q_k \right| \simeq 1$ is inserted into $\mathds{1} = (z\mathds{1}-\mathbf{H})\frac{1}{(z\mathds{1}-\mathbf{H})}$, $\left\langle q_m \left| z-\hat H \right| q_k \right\rangle$ is calculated using the Lanczos matrix, and $x_{k0}$ uses the continued fraction (\ref{contfrac}). Indeed, we reproduce this equality to approximately machine precision, confirming the completeness of the  Lanczos vectors in our calculations.

\section{Results and Discussions}

In this paper, we illustrate the SA-NCSM/GF method for the  $^4$He(n,n)$^4$He elastic scattering, which allows for comparisons to experiment and other \textit{ab initio} theoretical studies. The $^4$He target ground state, as well as the Lanczos algorithm for evaluating the Green's function in the $A+1$ and $A-1$ systems  through Eqs. (\ref{tiGFp_pT}) and (\ref{tiGFh}), respectively, are computed in the \textit{ab initio} SA-NCSM approach with the NNLO$_{\rm opt}$ NN chiral potential. This interaction minimizes the effect of the three body forces and has been shown to give an excellent description of nuclear structure and reaction observables; in addition, observables calculated with the NNLO$_{\rm opt}$ are found in a good agreement with those calculated with other chiral potentials that require the use of the corresponding three-nucleon forces  (see, e.g., Refs. \cite{Ruotsalainen19,WilliamsBCD2019, BurrowsEWLMNP19,BakerLBND20,Sargsyan_A8,PhysRevC.106.024001}).

We utilize SA-NCSM calculations for $\hw=12$-20 MeV in complete model spaces up to \Nmax=13 for $^{3,4,5}$He to accommodate both natural and unnatural parity, which in the case of the largest calculation implies total of 15 HO shells. The calculations become independent of $\hw$ for sufficiently large \Nmax~ model spaces, providing a parameter-free \textit{ab initio} prediction. For the Lawson procedure we use $\lambda_{\rm CM}=100$ MeV. Unless explicitly mentioned, all calculations utilize infinite-space ground-state energies $E_0^{\infty}$ that are extrapolated from the \Nmax=9, 11, 13 calculations using the Shanks extrapolation method \cite{DShanks55}, with uncertainties estimated based on variations in \hw~(while no experimental energies are used in the present evaluations, they can be straightforwardly used in the Green's function evaluation, if beneficial). More details on the energy extrapolation method are given in Sec. \ref{section:Nmax_hw}. 
\begin{figure}[t]
    \centering
    {\includegraphics[width=0.95\columnwidth]{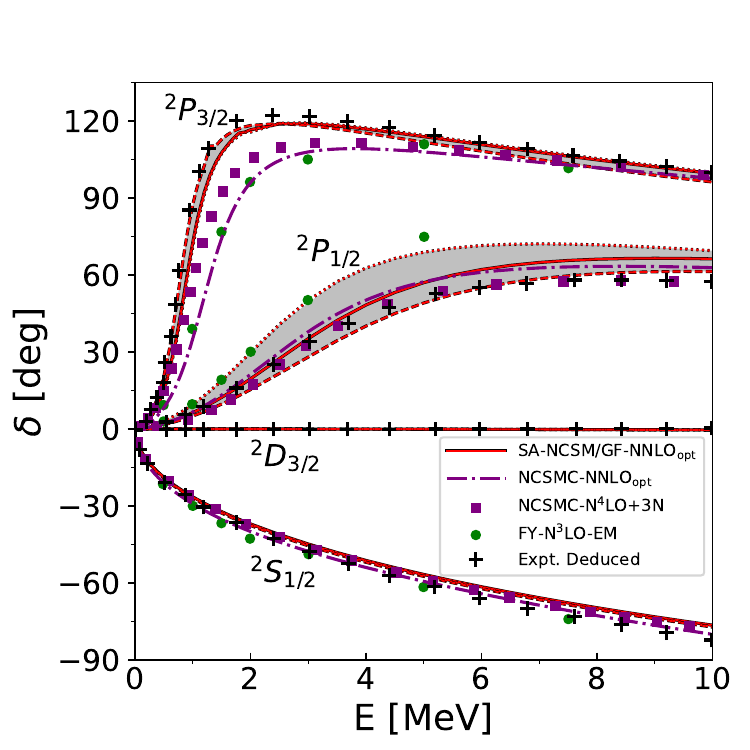}}
    \caption{Calculated n-$^{\rm 4}$He $^2S_{1/2}$, $^2P_{1/2}$, $^2P_{3/2}$, and $^2D_{3/2}$ phase shifts as a function of the energy in the CM frame using the \textit{ab initio} SA-NCSM/GF for \Nmax=13 across $\hbar\Omega=12$ (red dashed), 16 (red solid), and $20$ MeV (red dotted), with NNLO$_{\rm opt}$ NN interaction. These are compared to the NCSMC using multiple channels with $\hw=20$ MeV and the NNLO$_{\rm opt}$ NN interaction with $N_{\rm max}=17$ \cite{Zhang:2020rhz}  (purple squares), as well as a chiral N$^4$LO interaction including 3N forces  with $N_{\rm max}=11$ \cite{Kravvaris:2020lhp} (purple dotted dashed), the Faddeev-Yakubovsky (FY) approach \cite{Lazauskas2018} using the N$^3$LO-EM NN interaction \cite{EntemM03} (green circles), and the experimentally deduced values (black crosses) obtained using an $\mathbf{R}$-matrix analysis (see text for details).
    The gray bands show the $\hw=12$-20-MeV spread, to guide the eye. Two of the \hw~results are practically indistinguishable for the $P$ partial waves, and all \hw~results coincide for the $S$ and $D$ partial waves.
    }
    \label{fig:1}
\end{figure} 

For the phase shift results, we use the $\mathbf{R}$-matrix code of Ref.~\cite{Descouvemont16} with the SA-NCSM/GF  nonlocal optical potential $V_{J_0 \ell j}^J(r,r')$ of Eq. (\ref{optpot})  as input (see also Appendix \ref{section:Numerical}). 
We note that for $J_0=0$ of the target state, the potentials in the $(J_0 (\ell j))J$ coupling scheme  coincide with those used in the $(s \ell) J$ coupling scheme, $V_{\ell J}^{2s+1}(r,r')=V_{(J_0=0) \ell (j=J)}^J(r,r')$, where $s=\half$ is the channel spin, that is, the total spin of the $\alpha$ and neutron\footnote{
We use the $(s \ell) J$ scheme when we report phase shifts for $^{2s+1}\ell_J$ partial waves, to directly compare to earlier studies. All other quantities for $\alpha$+n are reported from the $(J_0 (\ell j))J$ coupling scheme for $\ell_J$ partial waves.
}.
\begin{figure}[t]
    \centering
    {\includegraphics[width=0.95\columnwidth]{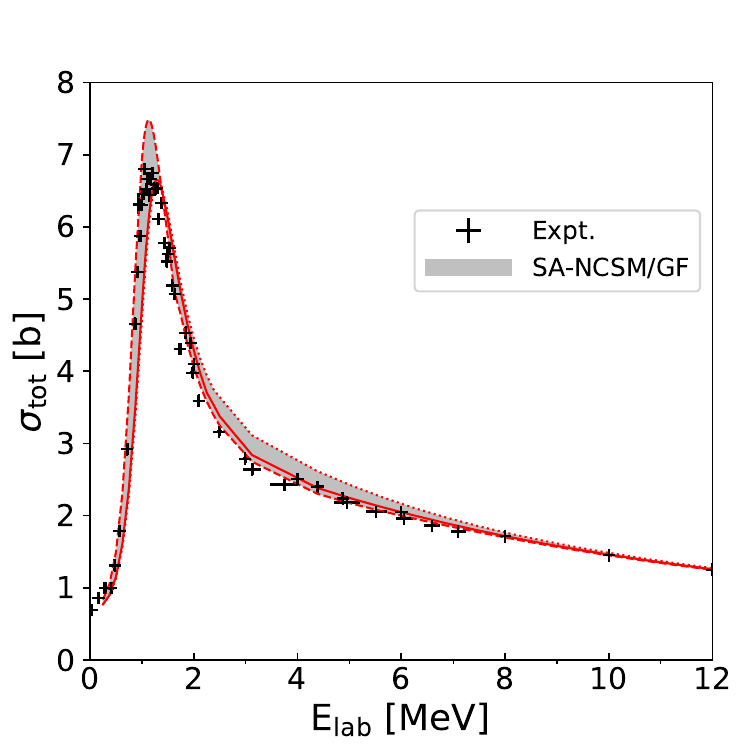}}
    \caption{
    Calculated n-$^{\rm 4}$He total cross sections vs. the laboratory-frame projectile kinetic energy obtained using the \textit{ab initio} SA-NCSM/GF across $\hbar\Omega=12$ (red dashed), 16 (red solid), and $20$ MeV (red dotted), and compared to three sets of experimental data \cite{Total_Cross_Section_Expt1, Total_Cross_Section_Expt2, Total_Cross_Section_Expt3} (labeled as ``Expt."). The gray band shows the $\hw=12$-$20$ MeV spread, to guide the eye. 
    There are energy ranges where curves are indistinguishable from each other.
    }
    \label{fig:2}
\end{figure}

The phase shift comparisons for n+$^4$He perform remarkably well when compared to both experimentally deduced values and earlier theoretical calculations (Fig. ~\ref{fig:1}). The experimentally deduced phase shifts are calculated from experimental total cross sections using an $\mathbf{R}$-matrix evaluation (see private communication G. M. Hale, Ref. [41] from \cite{Kravvaris:2020lhp}). 
We find that the $^2P_{\frac32}$ phase shifts from the SA-NCSM/GF, e.g., for \hw=16 MeV, yield a  threshold energy that agrees with the experimental one within 230 keV. This is important since reaction observables are very sensitive to the threshold energy. In addition, we
find a close agreement with the \textit{ab initio} many-body framework based on the RGM theory, the NCSMC \cite{BaroniNQ13}, where the microscopic structure of the clusters informs norm and Hamiltonian kernels, without an explicit construction of optical potentials. The NCSMC calculations use excitations of the target and two interactions: Refs. \cite{Zhang:2020rhz}  uses  the NNLO$_{\rm opt}$ NN interaction and Ref. \cite{Kravvaris:2020lhp}  uses a N$^4$LO chiral NN interaction with 3-body force. Interestingly, the two interactions yield very similar NCSMC results, whereas the SA-NCSM/GF with the NNLO$_{\rm opt}$ likely benefits from the use of infinite-space energies. Our results also agree with the Faddeev-Yakubovsky evaluation of the neutron  phase shifts \cite{Lazauskas2018}, which are close to the edge of the SA-NCSM/GF $\hw$ spread. This approach uses the N$^3$LO-EM NN interaction, and since it is practically exact, we expect that the differences in Fig.~\ref{fig:1} stem from the different interaction used. Similarly, the SA-NCSM/GF phase shifts agree with the ones derived in the single-state harmonic oscillator representation of scattering equations \cite{PhysRevC.98.044624} for different NN interactions. It will be interesting to compare all these methods for the NNLO$_{\rm opt}$ NN interaction and in the infinite-space limit.
\begin{figure*}[t]
    \centering
    {
    \includegraphics[width=0.32\textwidth]{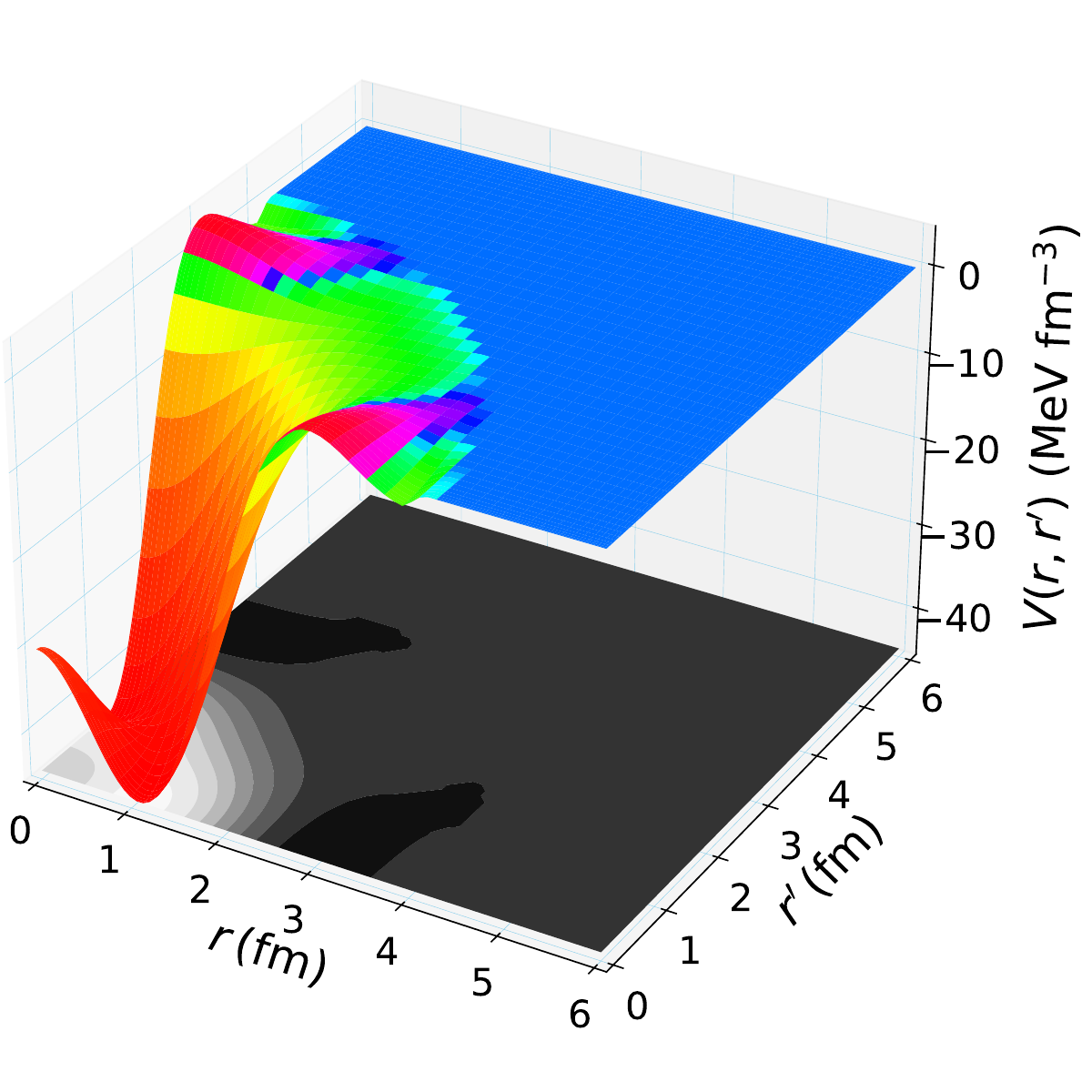}}
    {\includegraphics[width=0.32\textwidth]{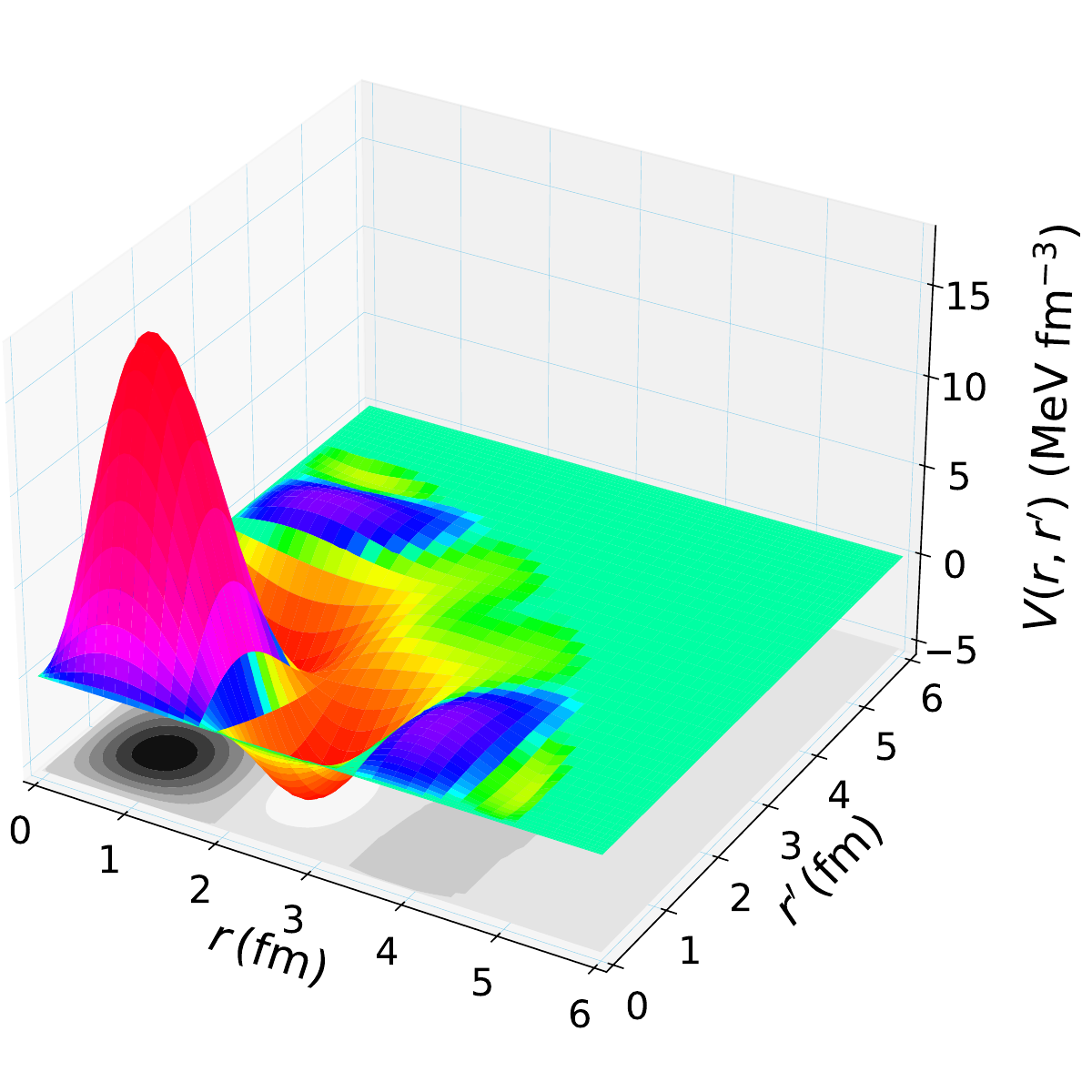}}
    {\includegraphics[width=0.32\textwidth]{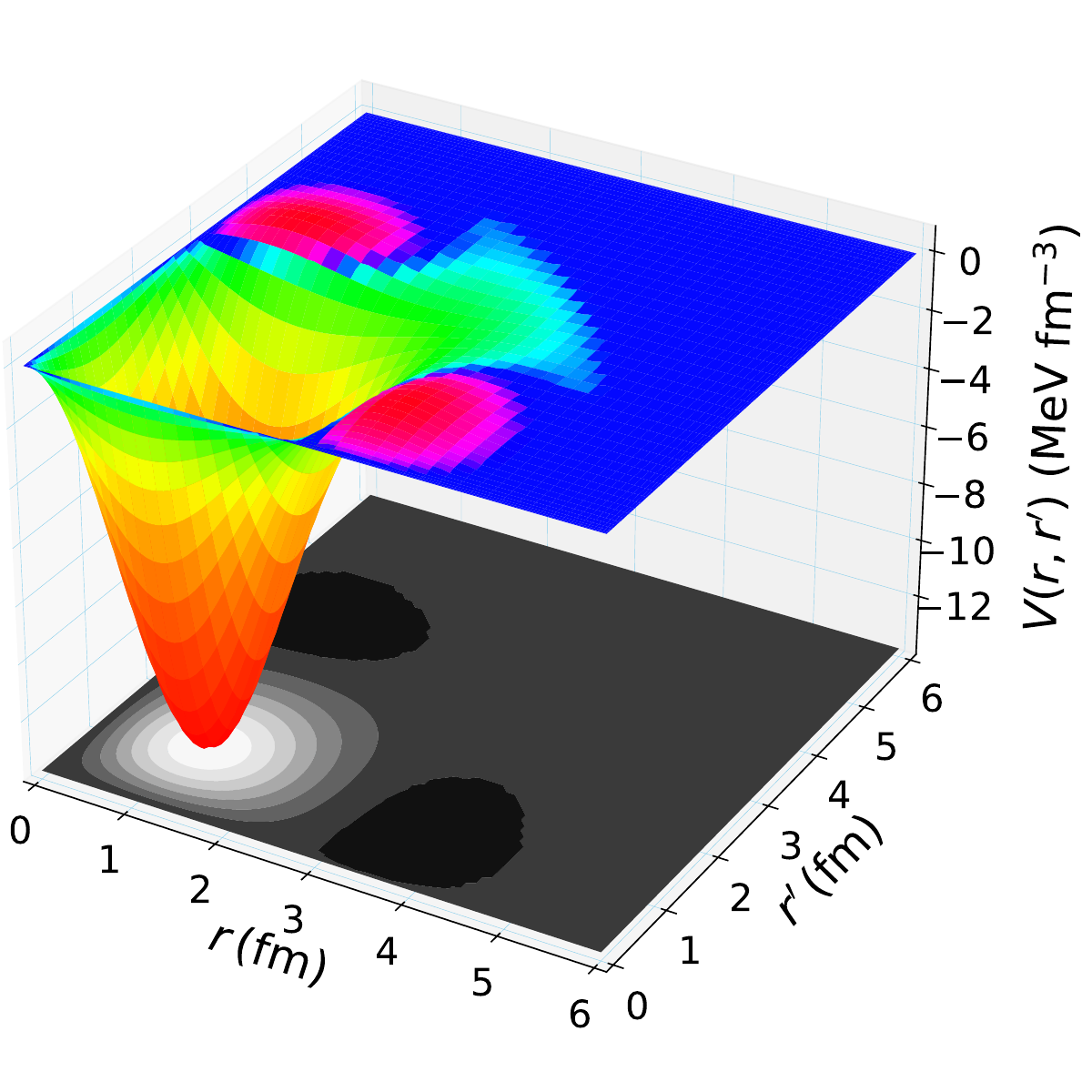}}\\
    (a) \hspace{2.1in } (b) \hspace{2.1in } (c) 
    \caption{The translationally invariant nonlocal n+$^4$He optical potential for the (a) $S_{\frac12}$, (b) $P_{\frac12}$, and (c) $P_{\frac32}$ partial waves, calculated in the \textit{ab initio} SA-NCSM/GF for $E=5.0$ MeV with $\epsilon=0$ MeV, $\hbar\Omega=16$ MeV, and $N_{\rm max}=13$.
}
    \label{fig:nonlocal_potentials}
\end{figure*}

From the phase shifts $\delta_{J_0 \ell j}^{J}(E)$, we evaluate the total cross section for elastic scattering (Fig. \ref{fig:2}) using,
\begin{eqnarray}
    \sigma_{\rm tot}(E) = \frac{4\pi}{k(E)^2} \sum_{\ell j J} \frac{(2J+1)\sin^2\delta_{J_0 \ell j}^{J}(E)}{(2 \half +1)(2J_{0}+1)},
    \label{eq:total_cross_section}
\end{eqnarray}
where $\half$ and $J_{0}$ in the denominator are the projectile and target total spins, respectively (and as mentioned above  $J_{0}$ is fixed by the reaction entrance channel), $k=\frac{\sqrt{2 \mu_p E}}{\hbar}$ is the wavenumber corresponding to the reaction energy $E$ in the CM frame \cite{ThompsonN09}. 
When comparing the total cross sections to experiment, we use the laboratory kinetic energy of the projectile,
$E_{\rm lab}=\frac{E}{\mu_p/m_N}=E\frac{(A+1)}{A}$.

Most importantly, the  
total cross sections for the SA-NCSM/GF calculations  agree remarkably well with experiment, as shown in Fig. \ref{fig:2} for projectile laboratory kinetic  energies $E_{\rm lab} \le 12$ MeV. As expected from the good description of the $^2P_{3/2}$ phase shifts, the cross-section peak energy is well reproduced in the SA-NCSM/GF approach. Notably, the spread of the calculated cross section arising from the \hw~variation is very small even though a significant \hw~range  is considered, whereas it further decreases across \hw=12-16 MeV while remaining in agreement with the data.

The \textit{ab initio} SA-NCSM/GF optical potentials $V_{\ell J}(r,r')$ (\ref{optpot}) that correspond to the  n+$^4$He  phase shifts discussed above are highly nonlocal  (Fig.~\ref{fig:nonlocal_potentials} for $E=5$ MeV). In general, they depend on the scattering energy $E$, however, for this system, we find almost no dependence for $E \le 12$ MeV for all $S_{\frac12}$, $P_{\frac12}$, $P_{\frac32}$, and $D_{\frac32}$ partial waves, except when $E$ is very close to a pole of the Green's function for $\epsilon=0$.
Although optical potentials are not observables and cannot be compared exactly between methods and inter-nucleon interactions used, similarities may exist in some features.
Interestingly, the neutron $S_{\frac12}$-wave potential (Fig.~\ref{fig:nonlocal_potentials}a) exhibits nonlocal peaks around 2.5 fm, attractive well at smaller distances, and an increase in strength at very small distances, which is similar to the neutron $S_{\frac12}$ partial wave for another closed-shell target of $^{16}$O when calculated with the NNLO$_{\rm opt}$ and \hw=20 MeV (see Fig. 7 in Ref.~\cite{RotureauDHNP17}). 
In addition, the potentials in Fig.~\ref{fig:nonlocal_potentials} should not be directly compared to the orthogonalized nonlocal potentials of Ref. \cite{QuaglioniN09} for the n-$\alpha$(g.s.), since the latter are calculated for each channel and several channels beyond the $\alpha$ ground state are used to obtain the NCSMC phase shift in Fig. \ref{fig:1}.  Nevertheless, there is similarity in the shape of the $P_{\frac12}$ partial wave from the RGM with N$^3$LO-EM NN interaction and \hw=19 MeV (Fig. 8 of Ref. \cite{QuaglioniN09}) and the one shown in Fig.~\ref{fig:nonlocal_potentials}b, albeit smaller in magnitude, whereas the optical potentials for the $S_{\frac12}$ partial wave is very different from the RGM n-$\alpha$(g.s.) effective interaction and allows for a bound state (see Appendix \ref{section:Numerical}).

\subsection{Spectral functions and imaginary contributions}
\label{section:Im}

Spectral functions are often used to study correlations and single-particle properties of the target nucleus, as they probe the probability density of removing a particle from a  single-particle state $\alpha$ in the target $J_0=0$ state at a given energy $E$ (with $E\le \varepsilon_F^-$):
\begin{eqnarray}
    S_h(\alpha,E) &=&  \sum_k \left| \lla \Psi_k^{A-1} \left| a_\alpha \right| \Psi_0^A \rra \right|^2  \delta\left(E - (E^A_0 - E_k^{A-1}) \right)  \nonumber \\
    &=& \frac{1}{\pi}{\rm Im}(G^{J-}_{(J_0=0)\alpha\alpha;E<\varepsilon_F^-}).
    \label{eq:Sh}
\end{eqnarray}
This means that spectral functions are readily available through the diagonal imaginary components of the Green's function [see Eq. (\ref{tiGFh_negE}]. This, in turn, defines the spectroscopic factor across all single-particle states that are occupied within the target state:
\begin{eqnarray}
    S^- &=& \sum_{\alpha} \int_{-\infty}^{\varepsilon_F^-} dE S_h(\alpha,E) = \sum_{\alpha} \mathcal{N}^h_{\alpha \alpha} \cr
    &=& \sum_{\alpha} \left(\frac{A}{A-1}\right)^{\frac{n_a+n_b}{2}} 
    \braket{\Phi^{J0-}_{\alpha}}{\Phi^{J0-}_{\alpha}}_{\rm L},
    \label{eq:Sfactor}
\end{eqnarray}
where we use Eq. (\ref{Eq.Hole_Norm}).
Similarly, using Eq. (\ref{Eq.Particle_Norm}), one can calculate the particle spectral function that describes the probability density for adding a particle at a single-particle state $\alpha$, $S_p(\alpha,E) = \sum_k \left| \lla \Psi_k^{A+1} \left| a_\alpha^\dagger \right| \Psi_0^A \rra \right|^2  \delta\left(E - (E_k^{A+1} - E^A_0 ) \right) $ = $-\frac{1}{\pi}{\rm Im}(G^{J+}_{(J_0=0)\alpha\alpha;E>\varepsilon_F^+})$ for a given energy $E$ ($ \ge \varepsilon_F^+$), with $S^+=\sum_{\alpha} \mathcal{N}^p_{\alpha \alpha}$.

\begin{figure}[!h]
    \centering
    {\includegraphics[width=0.9\columnwidth]{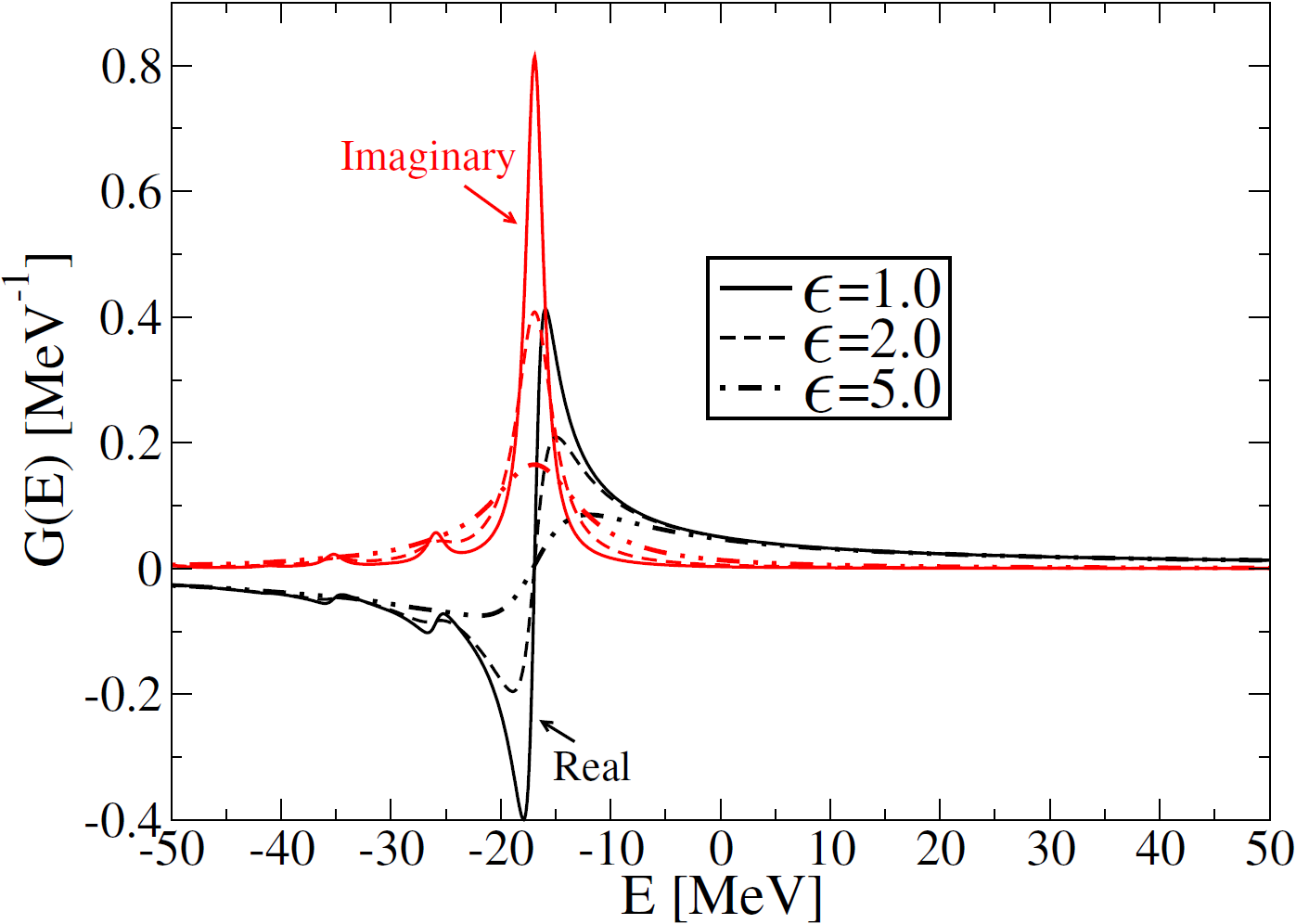}}
    \caption{Translationally invariant spectral functions $S_h$ for $^4$He ($\times \pi$), ${\rm Im} G(E,\epsilon)$, along with ${\rm Re} G(E,\epsilon)$   summed over $n_\alpha$ for the $s_{1/2}$ single-particle levels plotted against the energy in the CM frame, for $\epsilon= 1.0$ MeV (solid curves), 2.0 MeV (dashed curves), and  5.0 MeV (dotted dashed curves). Both the real (black) and imaginary (red) components of the Green's function are shown for each $\epsilon$ value.}
    \label{fig:spectral}
\end{figure}

Figure~\ref{fig:spectral} shows  the $\pi \sum_{n_\alpha}S_h(\alpha,E)$ spectral function for the $\ell_\alpha=0$ and $j_\alpha=1/2$ single-particle states across $\epsilon$ values of 1, 2, and 5 MeV, along with the real part of the Green's function. 
One can clearly recognize the $^3$He energies corresponding to  the peaks in the spectral functions (the lower the energy, the higher the excited states in $^3$He that contribute).  
In addition, these
spectral functions show a clear dependence upon $\epsilon$ in widening the peaks but the location of each peak remains unchanged. Clearly, the integral from $-\infty$ to $\varepsilon_F^{-}$ is independent of $\epsilon$ and equivalent to the spectroscopic factor $S^-$ of
 Eq.~(\ref{eq:Sfactor}) (practically, $\varepsilon_F^{+}$ is used for the upper limit of the integration for nonzero $\epsilon$ to accommodate the peak tail at $\varepsilon_F^{-}$, as discussed in Ref. \cite{DickhoffBook}). 
In this particular case we find for the $s_{1/2}$ single-particle levels $S^-_{\ell=0,j=\half}=0.897$ when using the trace of the norm of the hole states, or equivalently the overlap contribution, both of which can be calculated directly from the SA-NCSM wavefunctions, and $S^-_{\ell=0,j=\half}=0.885 \,($0.879$)$ when using the integrals for $\epsilon=1.0$ ($2.0$) MeV.

\begin{figure}[!h]
    \centering
    {\includegraphics[width=0.99\columnwidth]{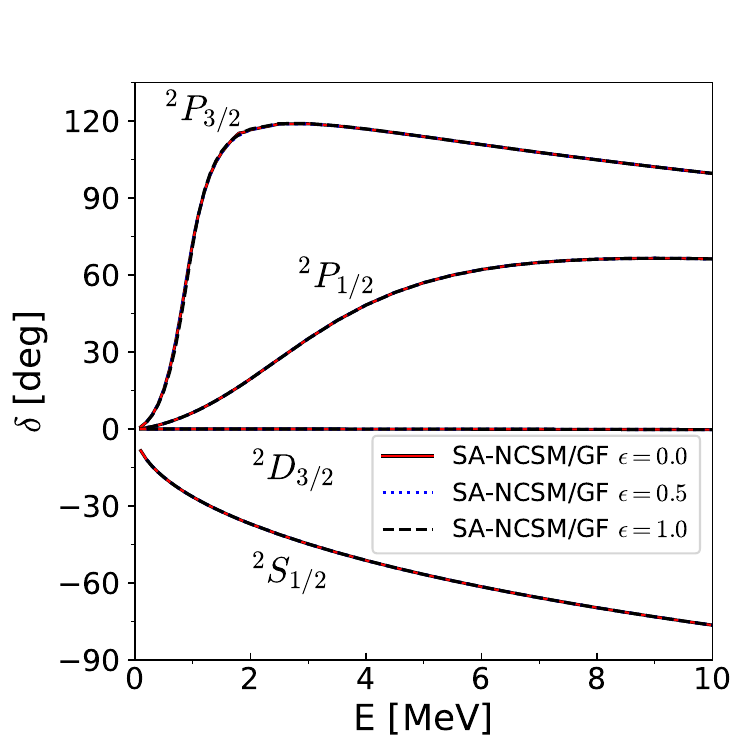}}
    \caption{A comparison of n-$^{\rm 4}$He $^2S_{1/2}$, $^2P_{1/2}$, $^2P_{3/2}$, and $^2D_{3/2}$ phase shifts with differing $\epsilon$ values (in units of MeV), for $N_{\rm max}=13$ and $\hw=16$ MeV (curves are indistinguishable).
    }
    \label{fig:eta_dependence}
\end{figure}

In addition, we study the dependence on $\epsilon$ and the imaginary contribution of the optical potential for the  $^2S_{\frac12}$, $^2P_{\frac12}$, and $^2P_{\frac32}$ phase shifts in n+$^4$He (Fig.~\ref{fig:eta_dependence}). It is very clear that $\epsilon$ has practically no effect on the (real) phase shifts calculated in the SA-NCSM/GF method. 
This is expected, since  in this energy regime only the neutron channel is open and there are two resonances only. Calculations can readily proceed for the Green's function parameter $\epsilon=0$ as far as the energy $E$ is slightly different from the $A+1$ SA-NCSM energies, that is, the poles in the Green's function. Analysis of the effects of $\epsilon$ on the absorption and its zero-limit impact, as discussed, e.g., in Refs. \cite{RAO1973182,SargsyanPKE23}, is left for a future study of the more intricate p+$^6$He system.

\subsection{Dependence on model-space parameters and infinite-space energies}
\label{section:Nmax_hw}

\begin{figure}[!h]
    \centering
    {\includegraphics[width=1\columnwidth]{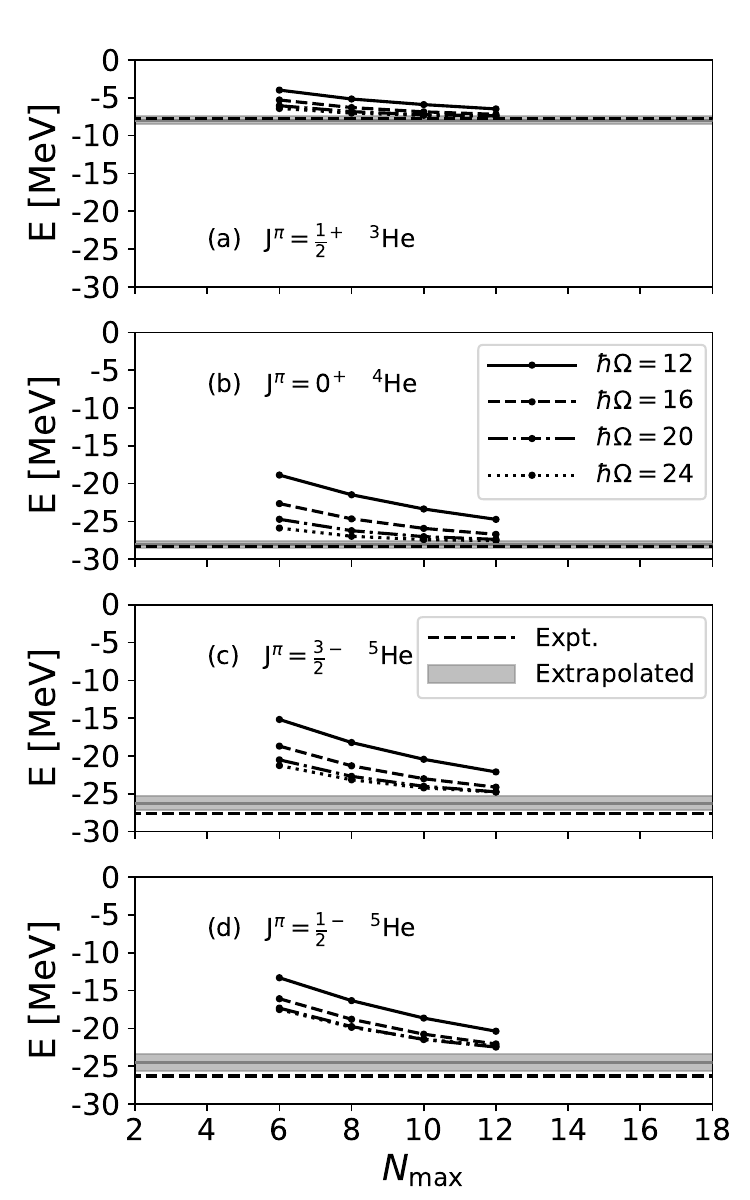}}
    \caption{Energy of (a) $\frac12^+$ g.s. of $^3$He, (b) $0^+$ g.s. of $^4$He, (c) $^5$He $\frac32^-$ g.s., and (d) the $^5$He $\half^-$ resonance with respect to $N_{\rm max}$ across several $\hw$ values (in units of MeV). The extrapolated values across all $\hw$ are given as a band with a centroid in the middle of the band, while the experimental energies are shown as a dashed line.}
    \label{fig:Energy_Convergence}
\end{figure}

To ensure \textit{ab initio} descriptions in no-core shell-model calculations, it is imperative to study convergence of results with respect to the $N_{\rm max}$ and $\hbar\Omega$ model-space parameters. We start with the structure calculations and examine energies of $^3$He, $^4$He, and $^5$He (see Fig.~\ref{fig:Energy_Convergence} for the $^3$He and $^4$He ground states and the  lowest-lying resonances in $^5$He). The g.s. and excitation energies of $^3$He and $^5$He are important for the description of n$+\alpha$, as they enter as poles in the Green's function. We emphasize that $^3$He lies energetically far away from $^4$He, namely, $\varepsilon^-_F=-20.6$ MeV, and while the $^3$He g.s. energy does not impact directly the n$+\alpha$ dynamics, it is critical for obtaining a bound state at this energy in the $S_{\frac12}$-wave optical potential. Since the SA-NCSM energies are on a converging trend with respect to $N_{\rm max}$, it is possible to extrapolate those to infinite-space energies in all $^{3,4,5}$He. Using Shank's extrapolation \cite{DShanks55}, applied and detailed in Refs. \cite{DytrychLDRWRBB20,BakerLBND20}, we determine the ground-state energy for $^3$He ($\frac12^+$ state), $^4$He ($0^+$ state) and $^5$He ($\frac32^-$ resonance), as well as the first excited $\frac12^-$ resonance in $^5$He, from \Nmax$=8$, 10, and 12 calculations, where uncertainties are estimated across an \hw=12-24 MeV range (the lowest SA-NCSM $^5$H $\frac12^+$ and $\frac32^+$ states are scattering states and their infinite-space energy is by default given by the threshold energy; see Ref. \cite{PhysRevC.98.044624} for convergence in the no-core shell model spaces). 
In Fig. \ref{fig:Energy_Convergence}, the centroid energies are shown as the mid-point within the \hw~region to guide the eye. 
In the SA-NCSM/GF evaluations we use the g.s. extrapolated energy for each \hw~ (for consistency with the wavefunction calculations). 
It is important to note that it is straightforward to use the infinite-space energies in the Lanczos algorithm, by simply substituting $z$ by $z^\infty=E  + E_0^{A,\infty} - (E_0^{A+1,\infty}-E_0^{A+1}) +\iu\epsilon$ for $G^+$ in the  continued fraction (\ref{contfrac}) (and similarly for $G^-$).

\begin{figure}[!h]
    \centering
    {\includegraphics[width=0.95\columnwidth]{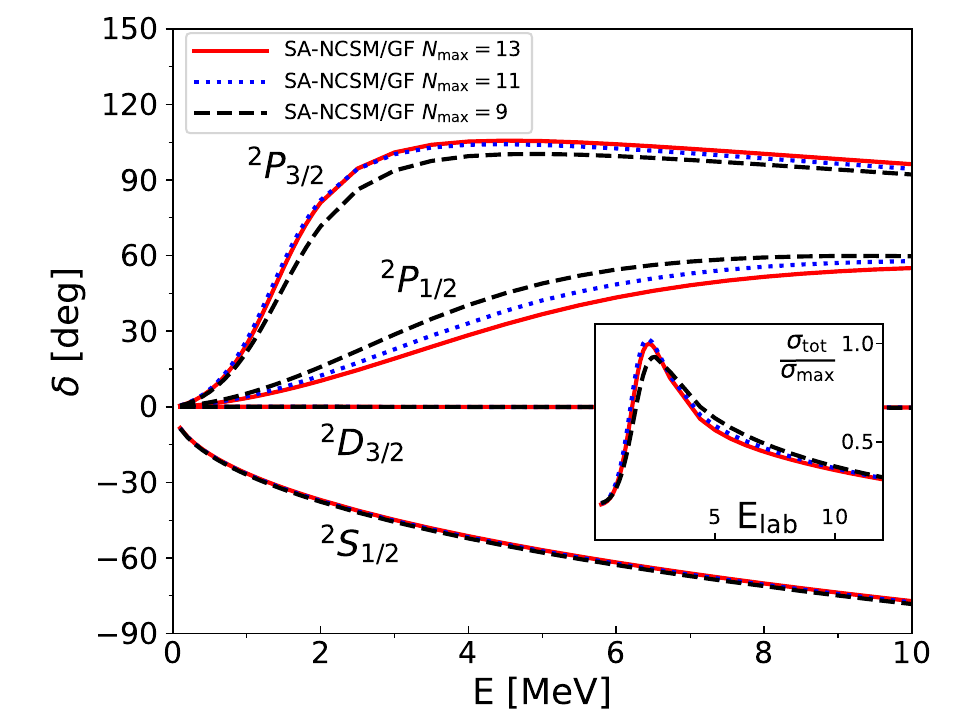}}
    \caption{A comparison of n-$^{\rm 4}$He $^2S_{1/2}$, $^2P_{1/2}$, $^2P_{3/2}$, and $^2D_{3/2}$ phase shifts with differing $N_{\rm max}$ values for $\hw= 16$ MeV and using the SA-NCSM g.s. energies. Inset: The corresponding total cross sections ($\sigma_{\rm tot}$) divided by the maximum total cross section ($\sigma_{\rm max}$) vs. laboratory energy $E_{\rm lab}$ (in units of MeV). There are no deviations for $^2D_{3/2}$ underneath the inset.}
    \label{fig:Phase_Shift_Nmax}
\end{figure}

Further, we study the dependence on \Nmax~ of 
the $^2S_{1/2}$, $^2P_{1/2}$, $^2P_{3/2}$, and $^2D_{3/2}$  phase shifts for n+$^4$He elastic scattering for a constant $\hbar\Omega=16$ MeV  using the SA-NCSM g.s. energies calculated in the \Nmax$=9$-$13$ model spaces (Fig. \ref{fig:Phase_Shift_Nmax}). 
This shows a clear convergence for all the partial waves, and especially a quick convergence for the $S$-wave or $D$-wave phase shifts, such that $N_{\rm max}=9$ or $11$ is already sufficient at capturing the correlations in the wavefunctions accessible at higher \Nmax~ values. In addition, the corresponding total cross sections  reflect the same converging trend (Fig. \ref{fig:Phase_Shift_Nmax}, inset). 
These outcomes corroborate the convergence of the resonance energies for $^2P_{1/2}$ and $^2P_{3/2}$ that are calculated as relative energies with respect to the threshold. Namely, it is interesting that -- different from the convergence of the absolute energies shown in Fig. \ref{fig:Energy_Convergence} -- the resonance energies are relatively stable in these $N_{\rm max}$ model spaces, with a very slow rate of decrease. Ultimately, their infinite-space extrapolation is needed to fully reproduce the resonant phase shifts  and the peak of the cross section.

\begin{figure*}[t]
    \centering
    {\includegraphics[width=0.32\textwidth]{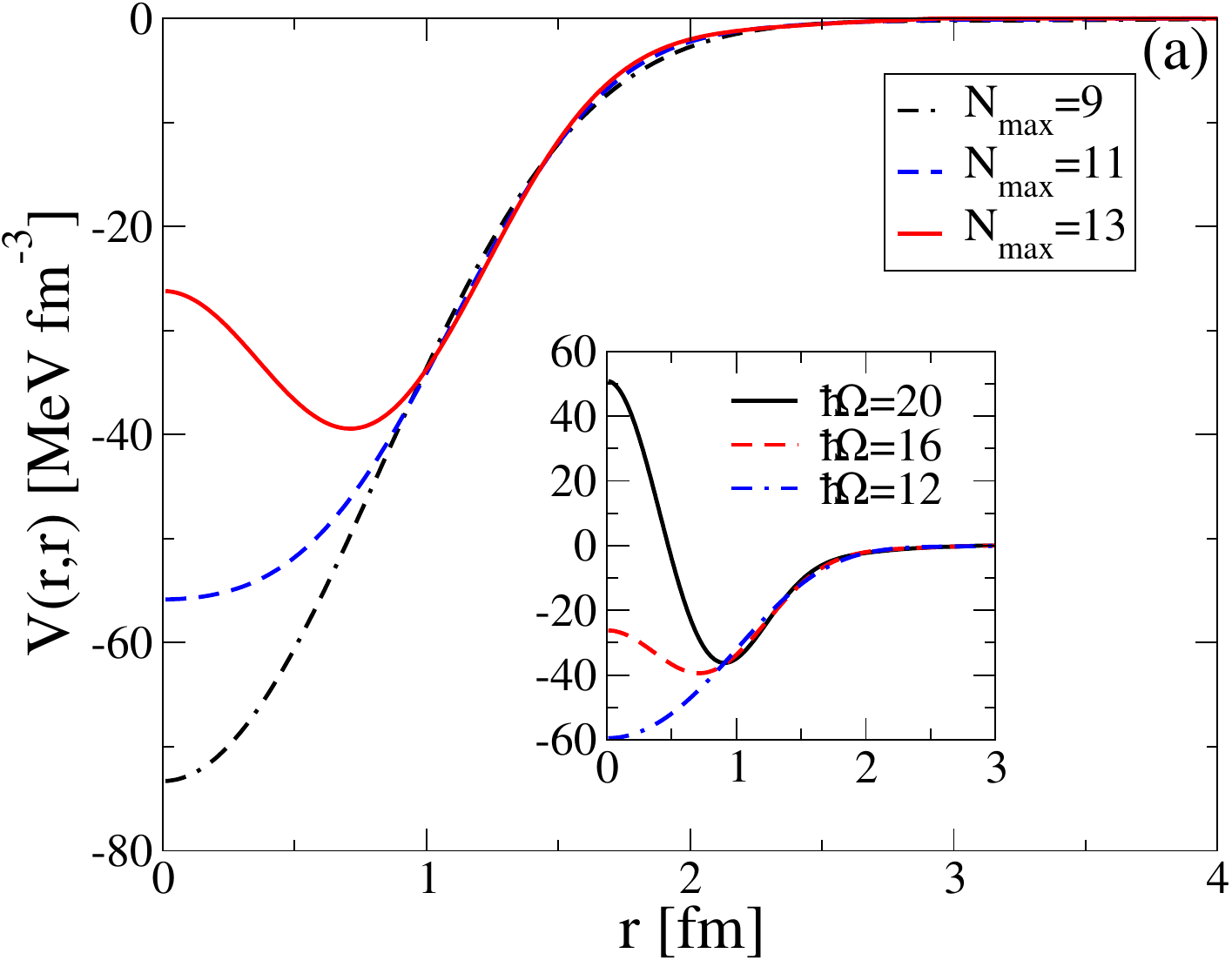}}
    {\includegraphics[width=0.32\textwidth]{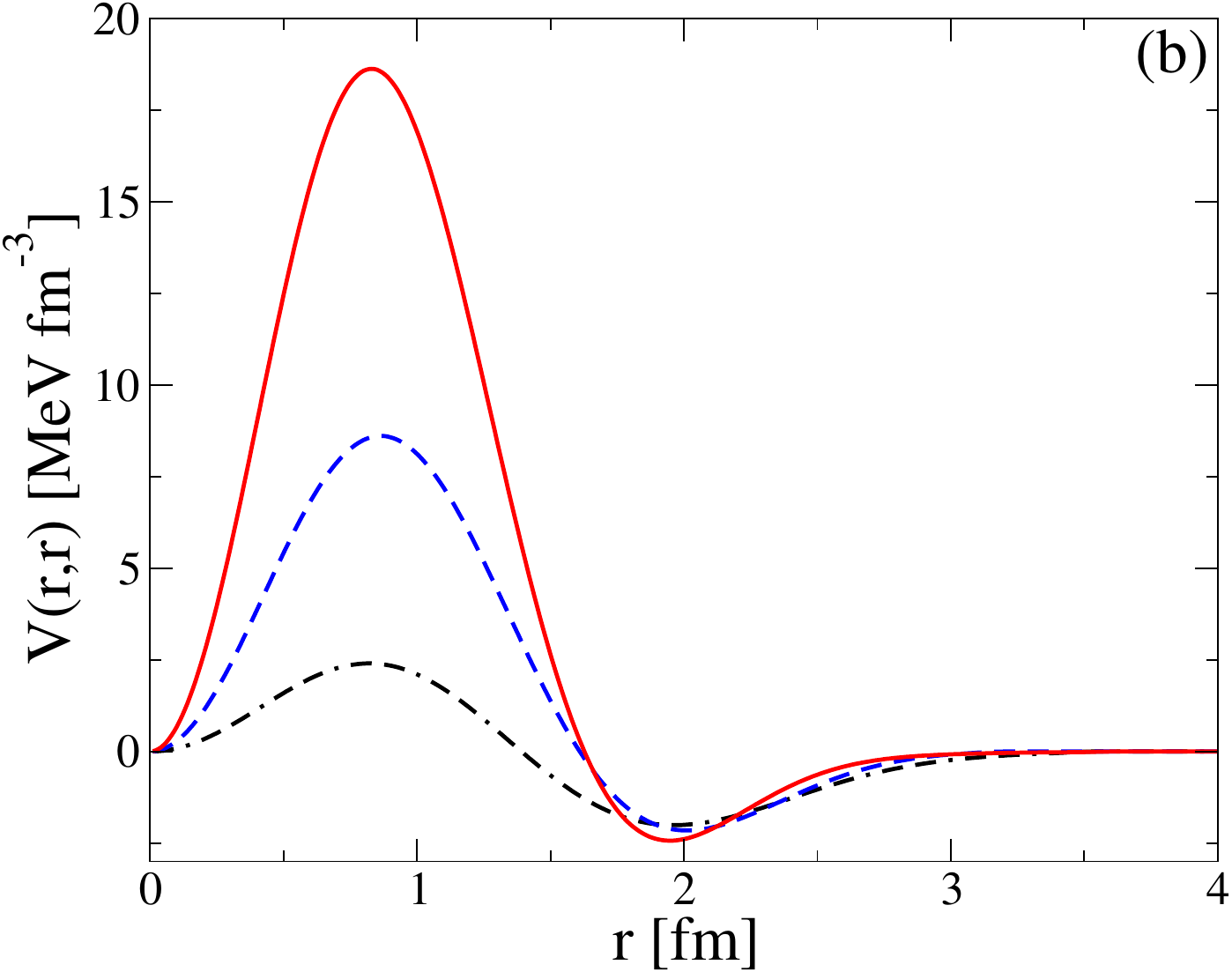}}
    {\includegraphics[width=0.32\textwidth]{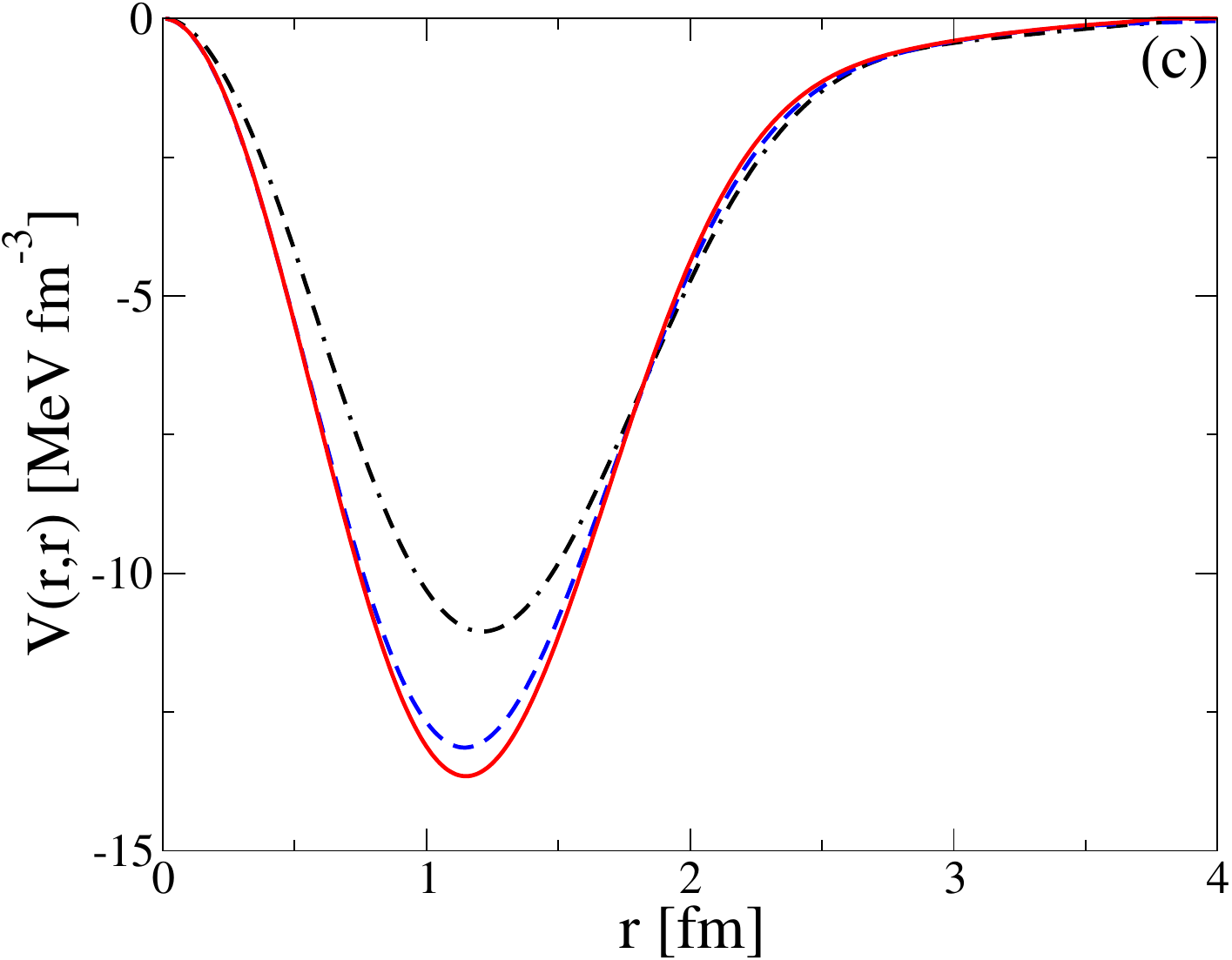}}
    \caption{A comparison of n-$^{\rm 4}$He (a) $S_{1/2}$, (b) $P_{1/2}$, and (c) $P_{3/2}$ translationally invariant diagonal potentials, $V_{\ell J}(r,r')$ with $r'=r$ (for the nonlocal counterparts for $\hw=16$ MeV, see Fig. \ref{fig:nonlocal_potentials}), with differing $N_{\rm max}$ values for $\hw=16$ MeV and $E=5.0$ MeV. Inset in (a): The same but for different $\hw$ values and fixed $N_{\rm max}=13$ for the $S_{1/2}$ partial wave.}
    \label{fig:Nmax_Comparison}
\end{figure*}

Figure~\ref{fig:Nmax_Comparison} shows a comparison of diagonal potentials, $V_{\ell J}(r,r'=r)$, across different $N_{\rm max}$ values. As we mentioned above, potentials are not observable, however, \textit{ab initio} deduced optical potentials may be used to inform phenomenological nonlocal optical potentials, as well as features related to short- and long-range correlations. As evident from Fig.~\ref{fig:Nmax_Comparison}, for each partial wave, the shapes of all the potentials are consistent, and importantly, they all agree near  the surface region (about $r \gtrsim 1.5$ fm) relevant to the energies in consideration.
The potentials differ significantly in magnitude in the interior region, which is only accessible at intermediate energies beyond the low-energy regime of the Green's function method considered. This is clearly seen in the $S$-wave potentials that coincide beyond $1.5$ fm, leading to the agreement in phase shifts shown in Fig. \ref{fig:Phase_Shift_Nmax}, whereas the higher resolution at larger \Nmax~ allows for the development of a ``repulsive core". Moreover, this feature becomes clearly enhanced and a repulsive core is observed at a higher \hw~value, which further improves the resolution of high-momentum phenomena (Fig. \ref{fig:Nmax_Comparison}a, inset). The strong dependence on \hw~in the interior region, from a Woods-Saxon-type potential at low \hw~ to a soft-core potential at high \hw, implies that these potentials should not be used at intermediate energies and beyond. For such energies, calculations in higher \Nmax~ are necessary to ensure that high-momentum components are properly treated and to achieve \hw~ independence.
In addition, the good agreement between $N_{\rm max}=11$ and $N_{\rm max}=13$ for the $P_{1/2}$ potential does not reflect in the phase shifts. This suggests a sensitivity to the nonlocal features that also affect the phase shifts.

\section{Conclusions}
In this paper, we have provided the first \textit{ab initio} translationally invariant optical potentials for the n+$^4$He $S_{1/2}$, $P_{1/2}$, $P_{3/2}$, and $D_{3/2}$ partial waves at low energies and show that they yield a remarkable reproduction of the experimental cross section for $^4$He(n,n)$^4$He elastic scattering. To construct these potentials, we have developed a novel SA-NCSM/GF approach to evaluate the single-particle time-ordered Green’s function that starts from realistic inter-nucleon interactions, and ensures a translational invariance by using the Lawson procedure, which is critical for light targets.
This provides effective nucleon-nucleus potentials that contain the information about all near reaction channels, including d and $\alpha$ partitioning.
This is achieved through the calculated $A\pm 1$ systems, which however makes the problem computationally intensive -- fortunately, solutions become feasible with the efficacious Lanczos algorithm and, in future studies, the use of selected model spaces for heavier nuclei. 

For n+$^4$He, the $^4$He target ground state, as well as the Lanczos algorithm for evaluating the Green's function in $^3$He and $^5$He are computed in the \textit{ab initio} SA-NCSM approach with the NNLO$_{\rm opt}$ chiral potential in complete model spaces up to 15 HO shells. We have shown that the n+$^4$He phase shifts calculated with infinite-space $^{3,4,5}$He ground-state energies agree with other \textit{ab initio} theoretical studies that provide $(A+1)$-body solutions without the explicit construction of optical potentials. 
Results suggest that both resonance energies and correlations play an important role in reproducing the $n+\alpha$ dynamics.
The SA-NCSM/GF yields a total cross section for the n+$^4$He elastic scattering that reproduces almost all of the data points within $1\sigma$ and is reasonably  independent of $\hw$, thereby providing a reliable parameter-free NA optical potential for energies up to $E \sim 12$ MeV. 

In addition, we have discussed the n+$^4$He \textit{ab initio}  optical potentials for the $S_{1/2}$, $P_{1/2}$, $P_{3/2}$, and $D_{3/2}$ partial waves that are nonlocal and interestingly, exhibit features that have been observed for heavier closed-shell nuclei. Remarkably, the GF approach that properly treats both the particle and hole sector is key to getting a deeply bound $\frac12^+$ state in the $S_{1/2}$ potential in addition to the $\frac12^+$ scattering state. Furthermore, the approach can readily provide spectral functions to probe correlations and single-particle properties of the target nucleus, as illustrated here for the $s_{1/2}$ single-particle levels.

The new developments provide a tool to construct \textit{ab initio} NA optical potentials for a broad range of nuclei accessible to the SA-NCSM, currently up through the Calcium region. These potentials can be used with any reaction model that can accommodate nonlocal potentials (the GF method is suitable to provide local approximations, as discussed in Ref. \cite{PhysRevC.66.034313}). Future work using this method will include proton elastic scattering, and elastic scattering of heavier nuclei, including $^{6}$He, $^{12}$C, $^{16}$O, and $^{40}$Ca. In addition, generalizations to deuteron elastic scattering by invoking two-nucleon overlaps for dA potentials and inelastic scattering are possible within this framework. Descriptions of deuteron breakup reactions in standard distorted-wave Born approximation methods can also use the parameter-free NA potentials produced in the SA-NCSM/GF approach, by utilizing first \textit{ab initio} proton-nucleus and neutron-nucleus potentials, along with the NN potential for the proton-neutron system, and ultimately including  the \textit{ab initio} dA deuteron-nucleus potentials when they become available.

\section{Acknowledgments}

We thank Michael Birse for useful discussions. This work was supported by the U.S. National Science Foundation (PHY-1913728, PHY-2209060), as well as in part by
the U.S. Department of Energy (DE-SC0023532, DE-FG02-93ER40756) and the Czech Science Foundation (22-14497S). 
This work was also supported in part under the auspices of the U.S. Department of Energy by Lawrence Livermore National Laboratory under Contract DE-AC52-07NA27344.
This work benefited from high performance computational resources provided by LSU (www.hpc.lsu.edu), the National Energy Research Scientific Computing Center (NERSC), a U.S. Department of Energy Office of Science User Facility at Lawrence Berkeley National Laboratory operated under Contract No. DE-AC02-05CH11231, as well as the Frontera computing project at the Texas Advanced Computing Center, made possible by National Science Foundation award OAC-1818253.

\appendix

\section{Properties of the Green's function and numerical precision}
\label{section:Numerical}

The SA-NCSM/GF effective potential $V(\mathbf{r},\mathbf{r'};E)$ provides the single-nucleon overlaps of the $A\pm 1$ wavefunctions and their energies, including the correct asymptotics, without the need for calculating those explicitly in the many-body approach of interest.
 In this section, we examine the potential and single-nucleon overlaps before using the $\mathbf{R}$-matrix method, and validate the results against explicit many-body wavefunction computations in the SA-NCSM framework. Indeed, for sufficiently large \Nmax~ model spaces, the interior part of the wavefunctions is accurately described by the SA-NCSM calculations. 
 
 For a given channel for $J_0=0$, the equation of motion for the single-nucleon overlap with $\beta=\{ n_\beta \ell_\beta j_\beta\}$ is given as \cite{PhysRevC.66.034313,Mahaux1991}: 
\begin{eqnarray}
   \sum_{n_\beta} \left[ \varepsilon_k^\pm \delta_{\alpha \beta} -(T_{\rm rel})_{{\alpha \beta}}-V^J_{\alpha\beta}(E) \right] \braket{\Phi^{J\pm}_{\beta}}{\Psi_k^{A\pm1}} 
   =0, 
    \label{eq:EoMovlp}
\end{eqnarray}
where $V^J_{\alpha\beta}(E)$ is calculated at $E$ close but not equal to $ \varepsilon_k^\pm$.
 Eq. (\ref{eq:EoMovlp}) is derived from the EoM of the s.p. propagator given in Eq. (\ref{eq:EoMGF}), or $(E\mathds{1}-\mathbf{T}_{\rm rel})\mathbf{G}-\mathbf{V} \mathbf{G}=\mathds{1}$ in configuration space, by using the Lehmann representation of the Green's function [see Eq. (\ref{GFcompl}) but for Jacobi coordinates], together with the completeness relation for the $J_0=0$ particle and hole states $\mathds{1}=\sum_k \braket{\Phi^{J+}_{\alpha}}{\Psi_k^{A+1}} \braket{\Psi_k^{A+1}}{\Phi^{J+}_{\beta}} +\braket{\Phi^{J-}_{\beta}}{\Psi_k^{A- 1}}\braket{\Psi_k^{A- 1}}{\Phi^{J-}_{\alpha}} = \mathcal{N}^p_{\alpha \beta}+\mathcal{N}^h_{\beta \alpha }$ [cf. Eq. (\ref{Eq:Norm})], and making use of the properties that $\varepsilon_k^\pm$ are non-degenerate and that the corresponding overlap functions are unique \cite{Mahaux1991}.
\begin{figure}[!h]
    \centering
(a)  $P_{1/2}$, $E=4.0$ MeV  \\
    {\includegraphics[width=0.4\textwidth]{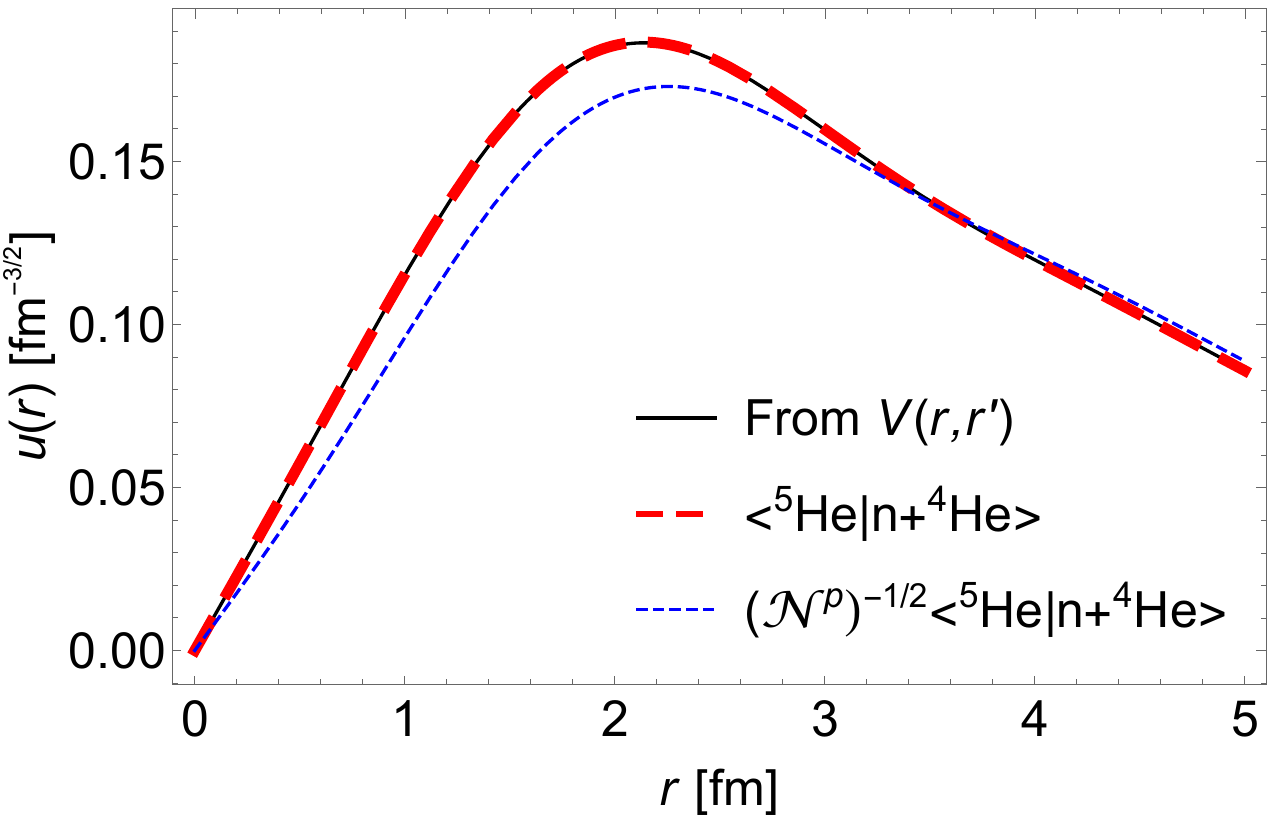}}\\
    (b) $P_{3/2}$, $E=1.8$ MeV  \\
     {\includegraphics[width=0.4\textwidth]{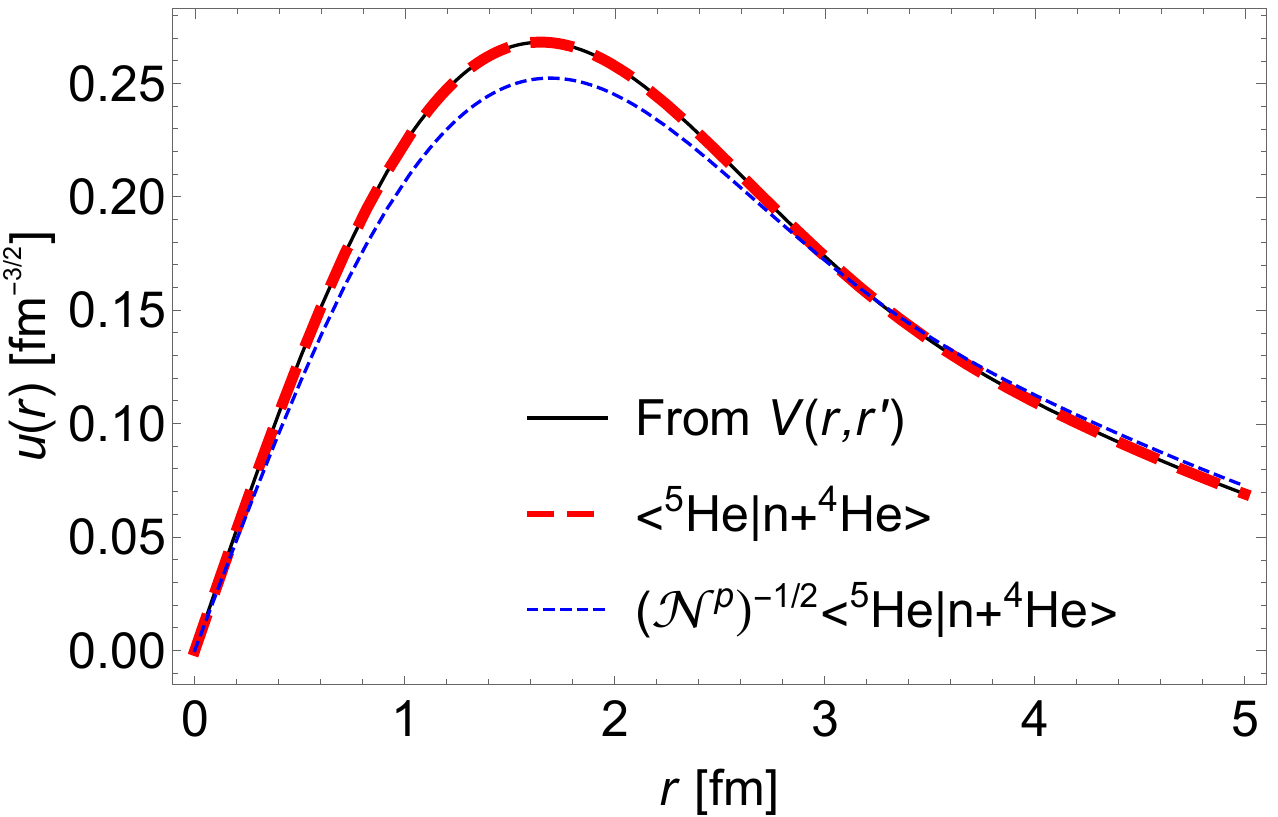}}      
    \caption{Eigenfunctions using the translationally invariant SA-NCSM/GF optical potential $V(r,r')$ for (a) $P_{1/2}$ at CM energy $E= 4.0$ MeV and (b) $P_{3/2}$ at CM energy $E= 1.8$ MeV, compared to the corresponding CM-free normalized overlap functions calculated in the SA-NCSM with spectroscopic factors $S=1.09$ $(1.15)$ for $P_{1/2}$ ($P_{3/2}$) (red long dashed), which are in addition modified within the particle-projected space, with spectroscopic factors $0.98$ $(0.98)$ for $P_{1/2}$ ($P_{3/2}$) (blue dashed). SA-NCSM calculations are for \Nmax=12 and \hw=16 MeV. 
}
    \label{fig:ovlp_cmp}
\end{figure}

Equation (\ref{eq:EoMovlp}) is a Schr\"odinger equation with $\mathbf{H}=\mathbf{T}_{\rm rel}+\mathbf{V} $, which yields eigenvectors that correspond to the normalized translationally invariant  overlaps $\frac{1}{\sqrt{S^\pm_{\ell j J; k}}}\braket{\Phi^{J\pm}_{\ell j}}{\Psi_k^{A\pm 1}} $ and eigenvalues that correspond to the associated $ \varepsilon_k^\pm$ energies; here, the spectroscopic factors for the specific channel and $A\pm 1$ state are defined as $S^\pm_{\ell j J; k}=\sum_{n}  | \braket{\Phi^{J\pm}_{n \ell j}}{\Psi_k^{A\pm 1}}| ^2$ and can be calculated through the energy derivative of the potential itself \cite{Mahaux1991}\footnote{
The effective potential can be used to solve an inhomogeneous equation with a source term, in which case the solutions are the overlaps functions and their norm provides the spectroscopic factor.
}. Indeed, with $\mathbf{V}=(E\mathds{1}-\mathbf{T}_{\rm rel})-\mathbf{G}^{-1}$, and hence $\mathbf{H} = E\mathds{1}-\mathbf{G}^{-1}$, the eigenvectors of $\mathbf{H}$ using the SA-NCSM/GF  evaluation of $\mathbf{G}^{-1}$ coincide exactly with the normalized translationally invariant  overlaps calculated within the SA-NCSM framework, as shown in Fig. \ref{fig:ovlp_cmp} for the $P_{1/2}$ and $P_{3/2}$ partial waves (similarly for the eigenvalues $\varepsilon_{J^\pi}=E^{A=5}_{J^\pi} -E^{A=4}_0$ for $J^\pi=\frac12^-$ and $\frac32^-$). We emphasize the importance of using an orthonormal basis in the particle-hole space (see the detailed discussion in Ref. \cite{PhysRevC.66.034313}). 
In contrast, some approaches, such as the cluster model (e.g, see \cite{LOVAS1998265,wildermutht77,rgm_tang_1978}), consider functions that are nearly complete in the space of particle (or hole) states only \cite{PhysRevC.66.034313},
which here corresponds to using $(\pmb{\mathcal{N}}^p)^{-1/2}\mathbf{G}^+(\pmb{\mathcal{N}}^p)^{-1/2}$. However, these overlap functions become modified by the norm of the non-orthonormal particle stats (see blue dashed curve in Fig. \ref{fig:ovlp_cmp}).  In addition, the particle-hole space is important for the $S_\half$ optical potential, e.g., when calculated at $E=0$ MeV (\Nmax=12 and \hw=16 MeV) it yields a bound state at $-20.8$ MeV that is occupied by the two protons and two neutrons of $^4$He in the $n+^4$He system. This is informed by the $^3$He $\frac12^+$ g.s. (at $-20.6$ MeV relative to the $^4$He g.s.), which enters as a pole in the hole-term of the Green's function.   

\begin{figure}[!h]
    \centering
    \includegraphics[width=0.41\textwidth]{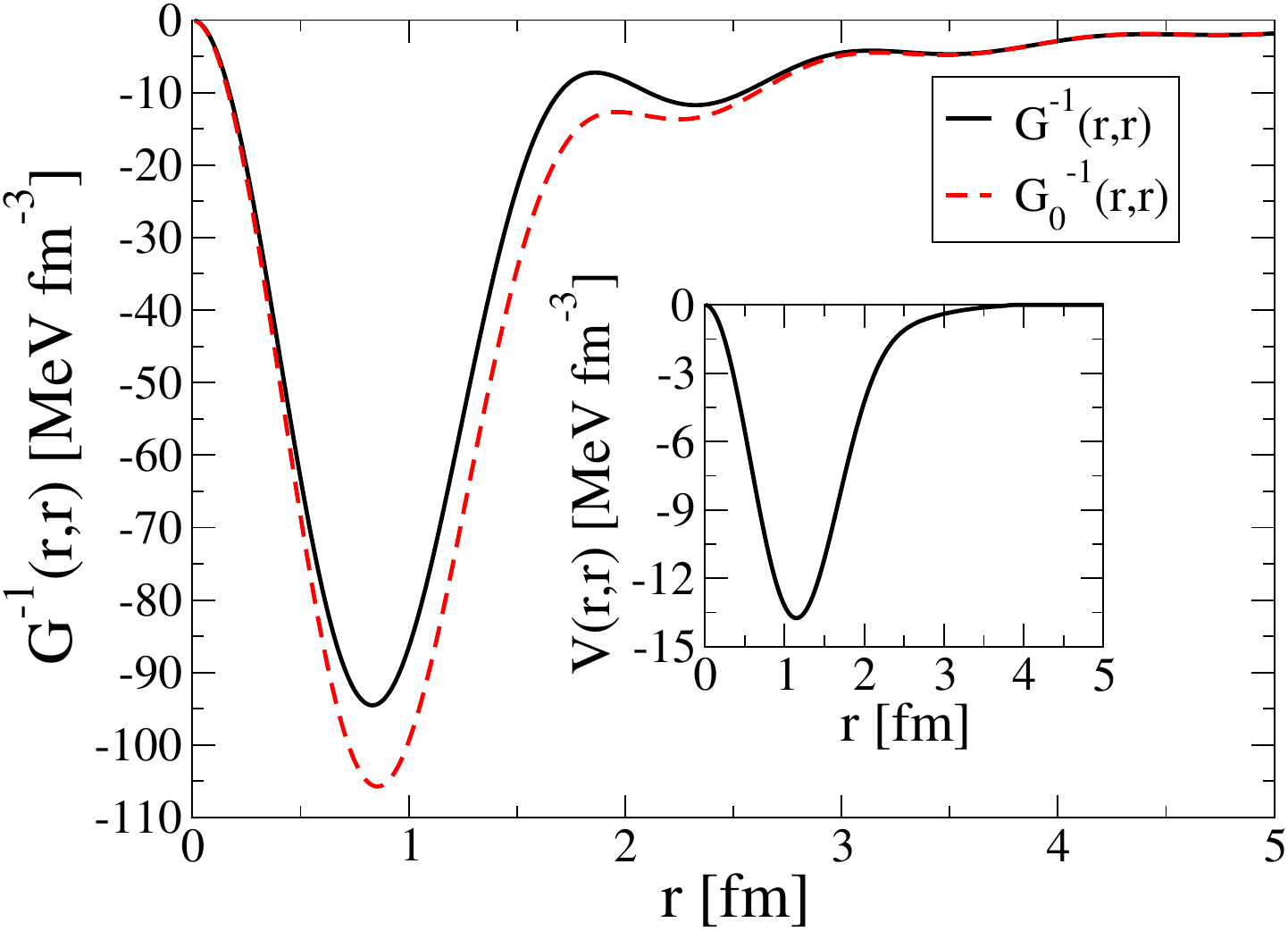}
    \caption{Translationally invariant $G^{-1}(r,r,E)$ and $G^{-1}_{0}(r,r,E)$ as a function of $r$ for the n$+^4$He $P_{3/2}$ partial wave for $N_{\rm max}=12$, $\hw=16$ MeV, and CM energy $E=5.0$ MeV. The inset shows the corresponding diagonal optical potential.}
    \label{fig:G_G0_Comparison}
\end{figure}

An important property of the effective potential derived in the SA-NCSM/GF framework is the finite interaction range. Indeed, beyond its effective range $R_0$, the neutron-target potential $\mathbf{V}=\mathbf{G}_0^{-1}-\mathbf{G}^{-1}=0$, and hence $\mathbf{G}_0^{-1}=\mathbf{G}^{-1}$, as seen in Fig. \ref{fig:G_G0_Comparison}. This makes calculations performed in a finite model space very suitable, as far as sufficiently high \Nmax~ values are used to accommodate the entire range up to $R_0$ and \hw~ independence is ensured for the observables at hand. This guarantees that the effective potentials are accurately described within the region of interest (for low-energy projectiles, this is typically $1.5$ fm $\lesssim r,r' \lesssim R_0$). However, because we perform the inversion of the Green's function in configuration space, one needs to take a special care of the matrix elements calculated for $n_r$ or $n_r'=n_r^{\max}$. For sufficiently large \Nmax~ model spaces, these matrix elements will contribute to long distances, where $\mathbf{G} \sim \mathbf{G}_0=(E\mathds{1}-\mathbf{T}_{\rm rel})^{-1}$. One can clearly see that inverting the tridiagonal matrix of the kinetic energy affects the $n_r^{\max}$ row and column of the inverted matrix due to the finite matrix sizes. To resolve this the sum in Eq. (\ref{optpot}) is taken to $n_{r}^{\max}-1$ (corresponding to $n_{\max}-2$). In addition, beyond $R_0$ the nodes of both $G^{-1}(r,r';E)$ and  $G_0^{-1}(r,r';E)$ coincide for $r > r'$ or $r < r'$, which helps address a  numerical precision at the sub-percent level in the calculations. Specifically, we slightly adjust, within 1\% or less, the HO characteristic length $b$ used in $G^{-1}(r,r';E)$ to ensure the node locations coincide with those of $G^{-1}_0(r,r';E)$. For example, for the $P_\frac12$ partial wave and $r=0.1$ fm, this means that a node at $r'=4.53$ fm appears now at $r'=4.49$ fm. Once the effective range of the interaction $R_0$ is identified through a region where $G^{-1}(r,r';E)$ and  $G_0^{-1}(r,r';E)$ coincide, we set the potential at larger distances exactly to zero (this is important since negligible numerical errors are enhanced at long distances in the $\mathbf{R}$-matrix method). To achieve a smooth transition to zero beyond $R_0$, we multiply $V(r,r';E)$ by a radial Gaussian function in the $r$ and $r'$ space with width $\sigma \sim 0.1$ fm. Varying the details of this procedure yields practically the same results with variation that is inconsequential compared to the \hw~ variation.

\section{Overlaps in laboratory and Jacobi coordinates}
\label{AppendixOverlaps}

For spatial coordinates, the transformation to Jacobi coordinates is (see, e.g., \cite{Navratil:2004tw}):
\begin{widetext}
\begin{eqnarray}
    u^{A+1*}_{ik,a ({\rm L})}
    &=&\sqrt{A+1} \int d \pmb{\xi}_1 \dots d\pmb{\xi}_{A-1} d\mathbf{R_{\rm CM}^A} d\mathbf{r}_{A+1}
      \Psi^{A+1}_{k\Gamma_k}(\pmb{\xi}_1 ,\dots, \pmb{\xi}_{A-1}, \mathbf{R_{\rm CM}^A}, \mathbf{r}_{A+1})^* \Psi^{A}_i(\pmb{\xi}_1 ,\dots, \pmb{\xi}_{A-1}) \phi_{\Gamma_i}(\mathbf{R_{\rm CM}^A})\phi_a(\mathbf{r}_{A+1}) \nonumber \\
      &=&\sqrt{A+1} \int d \pmb{\xi}_1 \dots d\pmb{\xi}_{A} d\pmb{\xi}_{0} 
      \Psi^{A+1}_k(\pmb{\xi}_1 ,\dots, \pmb{\xi}_{A})^* \phi_{\Gamma_k}(\pmb{\xi}_0)^* \Psi^{A}_i(\pmb{\xi}_1 ,\dots, \pmb{\xi}_{A-1})  \sum_{\Gamma \alpha} 
      \mathcal{M}^{\alpha+}_{ a; \Gamma_i \Gamma} 
       \phi_\alpha(\pmb{\xi}_{A}) \phi_{\Gamma}(\pmb{\xi}_0) \nonumber \\
      &=& \sum_{\alpha} \mathcal{M}^{\alpha+}_{ a; \Gamma_i \Gamma_k} \braket{ \Psi^{A+1}_k}{(\mathcal{A}  \Psi^{A}_i \phi_\alpha )  } 
      =\sum_{\alpha} \mathcal{M}^{\alpha+}_{ a; \Gamma_i \Gamma_k} \braket{\Psi_k^{A+1}}{\Phi^{+}_{i\alpha}},
      \label{ovlpp_TMB}
\end{eqnarray}
where $\mathcal{M}$ is defined in Eq. (\ref{TMB}) and we use $\int\phi^{*}_{\Gamma_k}(\pmb{\xi_0})\phi_{\Gamma}(\pmb{\xi_0}) d \pmb\xi_0 = \delta_{\Gamma \Gamma_k}$ for the CM of the $(A+1)$-body wavefunctions. Similarly, for the $A$ wavefunctions:
\begin{eqnarray}
    u^{A*}_{ki,a({\rm L})}
      &=&\sqrt{A} \int d \pmb{\xi}_1 \dots d\pmb{\xi}_{A-1} d\pmb{\xi}_{0}^A 
      \Psi^{A}_i(\pmb{\xi}_1 ,\dots, \pmb{\xi}_{A-1})^* \phi_{\Gamma_i}(\pmb{\xi}_0^A)^* \Psi^{A-1}_k(\pmb{\xi}_1 ,\dots, \pmb{\xi}_{A-2})  \sum_{\Gamma \alpha} 
      \mathcal{M}^{\alpha-}_{ a; \Gamma \Gamma_k} 
      \phi_{\Gamma}(\pmb{\xi}_0^A)\phi_\alpha(\pmb{\xi}_{A-1}) \nonumber \\
      &=& \sum_{\alpha} \mathcal{M}^{\alpha-}_{a; \Gamma_i \Gamma_k} \braket{ \Psi^{A}_i}{(\mathcal{A}  \Psi^{A-1}_k \phi_\alpha )  }=\sum_{\alpha} \mathcal{M}^{\alpha-}_{ a; \Gamma_i \Gamma_k}  \braket{\Psi_k^{A-1}}{\Phi^{-}_{i\alpha}}^*.
      \label{hovlprel2}
\end{eqnarray}

\end{widetext}
In this study, we use cluster basis that is free from CM excitations. In the $A+1$ CM reference frame, for the particle states (similarly for the hole states in the $A-1$ CM fame), one can relate the states projected onto the CM reference frame through  the  $\hat{\mathcal{P}}^0$ operator of Eq. (\ref{projCM}) to their t.i. counterparts  (true for any $\Gamma_i$, so w.l.g., we set $\Gamma_i=0$):
\begin{eqnarray}
   &&  \hat{\mathcal{P}}^0
    \left[\mathcal{A} \Psi^{A}_i(\pmb{\xi}_1 ,\dots, \pmb{\xi}_{A-1}) \phi_{\Gamma_i=0}(\mathbf{R_{\rm CM}^A})\phi_\alpha(\mathbf{r}_{A+1})\right] \cr
   && =  
   \sum_{\Gamma \beta} \mathcal{M}^{\beta +}_{ \alpha; 0 \Gamma} (\mathcal{A}\Psi^{A}_i(\pmb{\xi}_1 ,\dots, \pmb{\xi}_{A-1})  \phi_\beta(\pmb{\xi}_{A}) )\hat{\mathcal{P}}^0 \phi_{\Gamma}(\pmb{\xi}_0) \cr
   && =  
   \sum_{ \beta} \mathcal{M}^{\beta +}_{ \alpha; 0 0} (\mathcal{A}\Psi^{A}_i(\pmb{\xi}_1 ,\dots, \pmb{\xi}_{A-1})  \phi_\beta(\pmb{\xi}_{A}) ) \phi_{0}(\pmb{\xi}_0) \cr
    && = 
   \mathcal{M}^{\alpha+}_{ \alpha; 0 0} \mathcal{A}\Psi^{A}_i(\pmb{\xi}_1 ,\dots, \pmb{\xi}_{A-1})  \phi_\alpha(\pmb{\xi}_{A}) \phi_{0}(\pmb{\xi}_0).
\end{eqnarray}
Hence, 
\begin{eqnarray}
    \braket{\Psi^{A+1}_{k}}{\Phi^{+}_{i\alpha}} 
    =\frac{1}{\mathcal{M}^{\alpha+}_{ \alpha; 0 0}}
    \braket{\Psi^{A+1}_{k(\Gamma_k=0)}}{\Phi^{0+}_{i\Gamma_i=0\alpha}}_{\rm L}
    \label{projp}.
\end{eqnarray}

\section{Relations to RGM for the $J_0=0$ case} 
\label{appendix:B}

The connection to the RGM method and the RGM cluster basis states (see Refs. \cite{Wheeler_PR52_1937a} and its \textit{ab initio} realization in Ref. \cite{QuaglioniN09}) becomes clear when one examines the norm of the orthonormal basis used in the time-ordered Green's function in the particle-hole space for spatial dof and $J_0=0$.
According to Eq. (\ref{ovlpp_TMB}), the norm of the particle states in Jacobi coordinates is given as $\braket{\Phi^{+}_{\alpha}}{\Phi^{+}_{\beta}} 
      =\braketop{ \Psi^{A}_0 \phi_\alpha  } {\mathcal{A} \mathcal{A}} {  \Psi^{A}_0 \phi_\beta  } $, and hence
\begin{eqnarray}
     \braket{\Phi^{+}_{\alpha}}{\Phi^{+}_{\beta}} 
      &=& \braketop{\Psi^{A}_0 \phi_\alpha } {\mathds{1} } {  \Psi^{A}_0 \phi_\beta  } \cr
    &-&\braketop{ \Psi^{A}_0 \phi_\alpha  } {\sum_{i}^A\hat P_{i, A+1}} {  \Psi^{A}_0 \phi_\beta  } ,   
\end{eqnarray}
where $\hat P$ is a particle exchange operator.
Noting that $\braketop{\Psi^{A}_0 \phi_\alpha  } {\mathds{1} } { \Psi^{A}_0 \phi_\beta  }= \delta_{\alpha\beta}$, the norm of the particle and hole cluster basis states in the SA-NCSM/GF are related to the RGM norm $\mathcal{N}^{\rm RGM}_{\alpha\beta}$ [and associated  norm kernel $\mathcal{N}^{\rm RGM}_{\alpha\beta}(r,r')$] and the RGM exchange norm $\mathcal{N}^{\rm RGM,ex}_{\alpha\beta}$ as:
\begin{eqnarray}
     \mathcal{N}^p_{\alpha\beta}&=&\braket{\Phi^{+}_{\alpha}}{\Phi^{+}_{\beta}}=\braketop{ \Psi^{A}_0 \phi_\alpha  } {\mathcal{A} \mathcal{A}} {  \Psi^{A}_0 \phi_\beta  } =\mathcal{N}^{\rm RGM}_{\alpha\beta} \cr
     \mathcal{N}^h_{\beta\alpha}&=&\braket{\Phi^{-}_{\beta}}{\Phi^{-}_{\alpha}} 
     =\braketop{\Psi^{A}_0 \phi_\alpha  } {\sum_{i}^A\hat P_{i, A+1}}{  \Psi^{A}_0 \phi_\beta  } \cr
    &=&-\mathcal{N}^{\rm RGM,ex}_{\alpha\beta}.
    \label{linknormRGM}
\end{eqnarray}
These relations have been validated for the $E \ge \varepsilon_F^+$ regime, where the RGM is applicable. This further confirms the different technique used here based on CM-excitations-free cluster basis states that ensures  translationally invariant results. Importantly, for particle-target reaction processes, Eqs. (\ref{linknormRGM}) imply that while the antisymmetrization of the particle-target system is properly taken into account, first, the hole states take into account the exchange of the projectile with a nucleon in the target, and second, the orthonormal cluster basis states $\ket{\Phi_{\alpha}}$ used to evaluate the time-ordered Green's function describes \emph{non-antisymmetrized} states $\ket{\Psi^{A}_0 \phi_\alpha  }$. Indeed, using laboratory coordinates one can derive the $J_0=0$ norm of the hole states $\mathcal{N}^h_{ba}$ [that is, the density matrix elements for the target $\rho_{ba({\rm L})}$] starting from $-\mathcal{N}^{\rm RGM,ex}_{ab}$ in Eq. (\ref{linknormRGM}):
\begin{widetext}
\begin{eqnarray}
&& \braketop{\Psi^{A}_0 \phi_a  } {\sum_{i}^A\hat P_{i, A+1}}{  \Psi^{A}_0 \phi_b  }_{\rm L} 
    = A \int d\mathbf{r}_{1} \dots d\mathbf{r}_{A+1}  
    \braket{\Psi_0}{\mathbf{r}_{1},\dots, \mathbf{r}_{A}} \phi^*_{a}(\mathbf{r}_{A+1}) 
    \hat P_{A, A+1} \phi_{b}(\mathbf{r}_{A+1}) \braket{\mathbf{r}_{1},\dots, \mathbf{r}_{A}}{\Psi_0} \nonumber \\
    &&=A \int d\mathbf{r}_{1} \dots d\mathbf{r}_{A+1}  
    \braket{\Psi_0}{\mathbf{r}_{1},\dots, \mathbf{r}_{A}} \phi^*_{a}(\mathbf{r}_{A+1})  
    \hat P_{A, A+1} \frac{1}{\sqrt{A}}\sum_c \phi_{b}(\mathbf{r}_{A+1}) \phi_{c}(\mathbf{r}_{A}) \braket{\mathbf{r}_{1},\dots, \mathbf{r}_{A-1}}{a_c | \Psi_0}  \nonumber \\
    &&=A \int d\mathbf{r}_{1} \dots d\mathbf{r}_{A+1}  
    \braket{\Psi_0}{\mathbf{r}_{1},\dots, \mathbf{r}_{A}} \phi^*_{a}(\mathbf{r}_{A+1}) 
    \frac{1}{\sqrt{A}}\sum_c \phi_{b}(\mathbf{r}_{A}) \phi_{c}(\mathbf{r}_{A+1}) \braket{\mathbf{r}_{1},\dots, \mathbf{r}_{A-1}}{a_c | \Psi_0}\nonumber \\
    &&=\sum_{cd} \int d\mathbf{r}_{1} \dots d\mathbf{r}_{A}d\mathbf{r}_{A+1} 
    \braket{\Psi_0 | a^\dagger_d}{\mathbf{r}_{1},\dots, \mathbf{r}_{A-1}}
    \phi^*_{d}(\mathbf{r}_{A})\phi^*_{a}(\mathbf{r}_{A+1})
     \phi_{b}(\mathbf{r}_{A}) \phi_{c}(\mathbf{r}_{A+1}) \braket{\mathbf{r}_{1},\dots, \mathbf{r}_{A-1}}{a_c | \Psi_0} \nonumber \\
    &&= \braket{\Psi_0 | a^\dagger_b  a_a }{\Psi_0}_{\rm L} =\rho_{ba({\rm L})} = \braket{\Phi^{-}_{b}}{\Phi^{-}_{a}}_{\rm L}.
\end{eqnarray}

\end{widetext}

\bibliography{sancsmgf}

\end{document}